\documentclass[aps,prl,final,twocolumn,superscriptaddress]{revtex4-1}

\usepackage{amsmath}
\usepackage{braket}
\usepackage{graphicx}
\usepackage{epsfig}
\usepackage{natbib}
\usepackage{color}

\newcommand{\CrNbS}{$\mathrm{CrNb}_3\mathrm{S}_6$}

\newcommand{\hn}{\hat{n}}
\newcommand{\hx}{\hat{x}}
\newcommand{\hy}{\hat{y}}
\newcommand{\hz}{\hat{z}}

\begin{document}


\title{Dynamics of chiral solitons driven by polarized currents in monoaxial helimagnets}


\author{Victor Laliena}
\email[]{laliena@unizar.es}
\affiliation{Aragon Material Science Institut 
  (CSIC -- University of Zaragoza) and Condensed Physics Matter Department, University of Zaragoza \\
  C/Pedro Cerbuna 12, 50009 Zaragoza, Spain}
\author{Sebastian Bustingorry}
\email[]{sbusting@cab.cnea.gov.ar}
\affiliation{
  Instituto de Nanociencia y Nanotecnolog\'{\i}a, CNEA-CONICET, Centro At\'omico Bariloche,
  (R8402AGP), S. C. de Bariloche, R\'{\i}o Negro, Argentina}
\author{Javier Campo}
\email[]{javier.campo@csic.es}
\affiliation{Aragon Material Science Institut 
  (CSIC -- University of Zaragoza) and Condensed Physics Matter Department, University of Zaragoza \\
  C/Pedro Cerbuna 12, 50009 Zaragoza, Spain}


\date{July 14, 2020}

\begin{abstract}
Chiral solitons are one dimensional localized magnetic structures that are metastable in some ferromagnetic systems
with Dzyaloshinskii-Moriya interactions and/or uniaxial magnetic anisotropy.
Though topological textures in general provide a very interesting playground for new spintronics phenomena, how to properly create and control single chiral solitons is still unclear.
We show here that chiral solitons in monoaxial helimagnets, characterized by a uniaxial Dzyaloshinskii-Moriya interaction, can be stabilized with external magnetic fields.
Once created, the soliton moves steadily in response to a polarized electric current, provided the induced spin-transfer torque has a dissipative (nonadiabatic) component. 
The structure of the soliton depends on the applied current density in such a way that steady motion exists only if the applied current density is lower than a critical value,
beyond which the soliton is no longer stable.
\end{abstract}

\pacs{111222-k}

\maketitle


Magnetic structures of nanometric size, like domain walls, vortices, or skyrmions, attracted great attention since they are very
promising as the building blocks of spintronic components such as memories, logical gates, etc. 
To be useful, they have to satisfy at least two essential requirements: 1) be (meta)stable, and 2) move in a controlled way under the action
of external stimuli, such as applied magnetic fields or electric currents. 
Comparatively, chiral solitons have received much less attention, although they are also potentially useful in spintronics and digitalization applications.
These are solitonic magnetic structures of topological nature.
In monoaxial helimagnets, the Dzyaloshinskii-Moriya interaction (DMI) and the uniaxial magnetic anisotropy (UMA) are the key ingredients
that provide the soliton (meta)stability. At low enough temperatures and applied magnetic field the chiral solitons condense and form a chiral soliton
lattice (CSL) \cite{Dzyal64,Izyumov84,Kishine05,Togawa12,Laliena16a,Laliena16b,Laliena17a,Laliena18a}.
Solitons condense also in some regions of the phase diagram of cubic helimagnets, in the form of
skyrmion lattices~\cite{Muehlbauer09,Laliena17b,Laliena18b}.

As a new route to spintronic devices, chiral solitons may have advantages over skyrmions, whose motion is girotropic and therefore difficult to
control~\cite{Nagaosa13, Jiang16, Zhou19}, and over domain walls, since chiral solitons may provide a different route to avoid pinning effects hindering
domain wall motion~\cite{Yang15, Jeudy18, HerreraDiez19}.
As we will show here, chiral solitons in monoaxial helimagnets move steadily under the application of a polarized current, reaching velocities
of the order of $100\,\mathrm{m/s}$ for currents around $100\,\mathrm{GA/m^2}$.
Furthermore, if the current is large enough the stability of the soliton is compromised and the system is forced to a homogeneous magnetization state.

\begin{figure}[t!]
\centering
\includegraphics[width=0.4\textwidth]{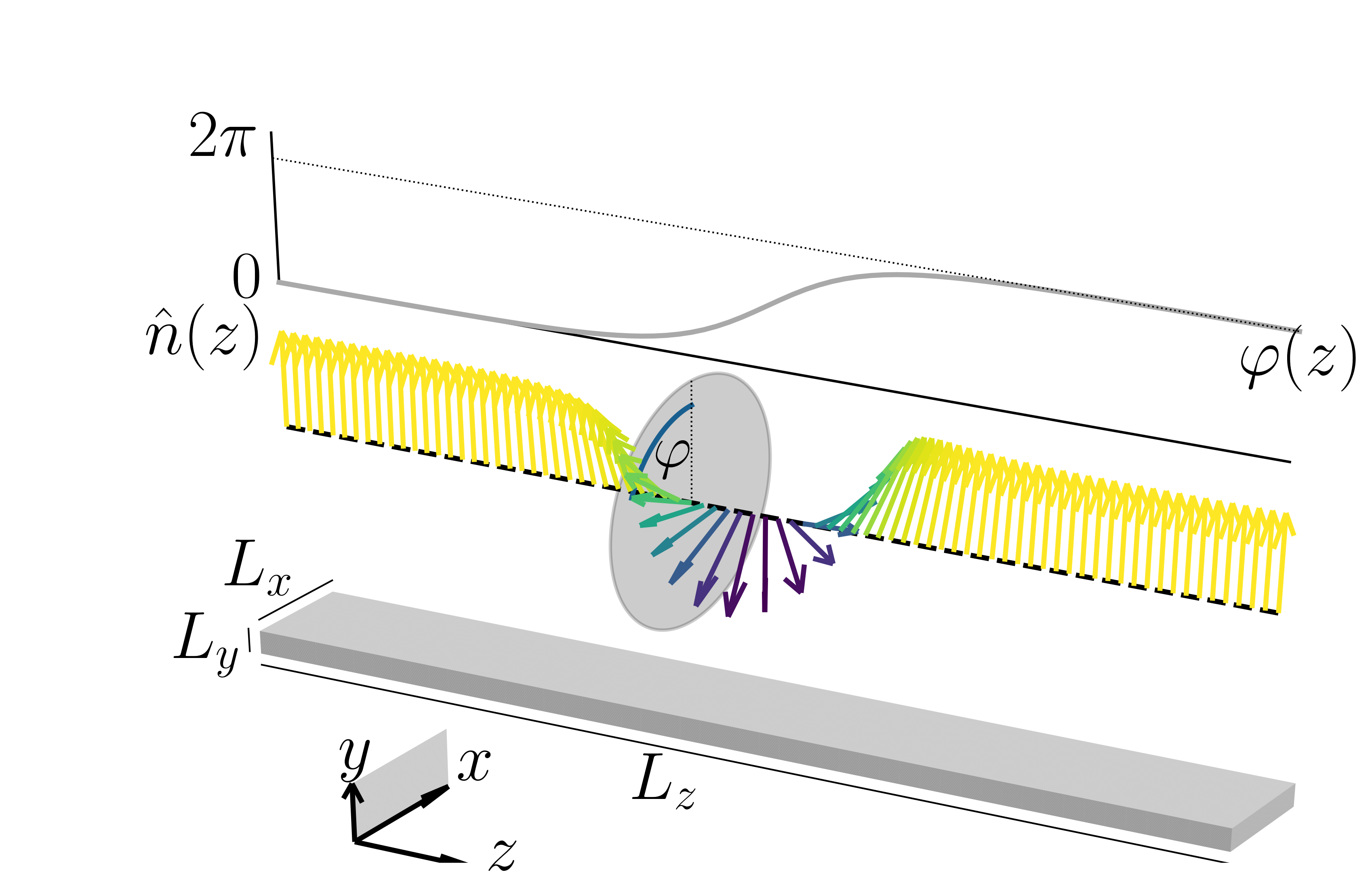}
\caption{
  Scheme of the chiral soliton ($\chi = +1$) described by $\hn(z)$ with $\varphi(z)$ given by $\varphi_0$ in Eq.~(\ref{eq:phi0}) (top).
  The polar angle is $\theta = \pi/2$ and thus the normalized magnetization $\hn$ is in the $x$-$y$ plane and rotates along the chiral axis,
  as indicated. The dimensions of the modeled magnetic track is schematically shown in the bottom figure.
\label{fig:1}}
\end{figure}

Consider a magnetic nanometer size track with dimensions \mbox{$L_y\ll L_x \ll L_z$} (see Fig.~\ref{fig:1}),
made of a monoaxial helimagnet, such as \CrNbS, with chiral axis along $\hz$. Its magnetic energy is given by $E = \int d^3x W$, with 
\begin{equation}
  W \!\!=\! A\!\!\!\!\sum_{i=x,y,z}\!\!\!\!\partial_i\hn\cdot\partial_i\hn-D\hz\cdot(\hn\times\partial_z\hn)
  -K(\hz\cdot\hn)^2-M_\mathrm{S}\vec{B}\cdot\hn, \label{eq:E}
\end{equation}
where $\hn$ is a unit vector field that describes the magnetization direction at each point of the film, $A$, $D$, and $K$ stand for the exchange
stiffnes constant, and the DMI and UMA strength constants, respectively, $M_\mathrm{S}$ is the saturation magnetization, and $\vec{B}$ is the
applied magnetic field. The DMI acts only along the $\hz$ axis, defining thus a monoaxial helimagnet, and it is of bulk type and not interfacial,
in spite that the track lies in a thin film in the $y=0$ plane.
The sign of $D$ is reversed if we reverse the direction of the $\hz$ axis, so that, with no loss of generality, we take $D>0$.
It is also convenient to introduce $q_0=D/2A$, which has the dimensions of inverse length, and the dimensionless parameters $\kappa=4AK/D^2$
and $\vec{h}=(2AM_\mathrm{S}/D^2)\vec{B}$. For the sake of simplicity, we ignore the magnetostatic energy, whose main effect could be approximately
taken into account by introducing magnetic anisotropies in the $(x,y)$ plane.
We do not expect it would change the qualitative conclusions of this work.

The effective field acting on the vector field $\hn$ is
\begin{equation}
\vec{B}_{\mathrm{eff}} \!= \!\frac{2A}{M_\mathrm{S}} \!\left[ \nabla^2\hn - 2q_0\hz\times\partial_z\hn+ q_0^2\kappa(\hz\cdot\hn)\hz + q_0^2\vec{h} \right].
\end{equation}
The corresponding Euler-Lagrange equations describing the static solutions are $\vec{B }_{\mathrm{eff}} = \lambda\hn$,
with $\lambda$ a Lagrange multiplier enforcing the constraint $\hn^2=1$. 
For $\vec{h}=0$ the ground state is a helical structure propagating along the $\hz$ axis with wave number $q_0$.
By applying a field parallel to the chiral axis, the helical state becomes a conical state, while if the applied field is perpendicular to the chiral axis a CSL
is formed \cite{Dzyal64,Izyumov84}. If the field is large enough, a transition to the homogeneous ferromagnetic (FM) state takes place,
whose nature depends on the angle between the magnetic field and the chiral axis~\cite{Laliena16a,Laliena16b,Laliena17a,Laliena18a}.

In the case of an applied field perpendicular to the chiral axis, the system has a single chiral soliton as a metastable static solution.
Taking $\vec{h} = h_y\hy$ and using the parametrization
\begin{equation}
\hn = -\sin\theta\sin\varphi\hx + \sin\theta\cos\varphi\hy +
\cos\theta\hz, \label{eq:hn}
\end{equation}
we seek for a solution of the Euler-Lagrange equations independent of $x$ and with constant $\theta$.
We obtain $\theta=\pi/2$ and the Sine-Gordon equation 
$\varphi^{\prime\prime}=q_0^2h_y\sin\varphi$, where the prime stands for the derivative with respect to $z$.
The solutions, $\varphi_0$, that have boundary conditions (BC) $\varphi_0(z=-\infty)=0$ and
$\varphi_0(z=+\infty) = \chi 2\pi$,
with $\chi=\pm 1$, are chiral solitons with helicities $\chi$, given by
\begin{equation}
\varphi_0(z) = 4 \chi \arctan\mathrm{e}^{q_0\sqrt{h_y}z}, \label{eq:phi0}
\end{equation}
The soliton width, $\Delta_0=1/q_0\sqrt{h_y} = \sqrt{2A/M_\mathrm{S}}/B_y$, is independent of $K$ and $D$
and it is controlled by the applied magnetic field.
The generic shape of the soliton with helicity $\chi=+1$ is shown in Fig.~\ref{fig:1}.

The soliton exists as a stationary point of the energy even in a simple ferromagnet. Whilst the DMI and UMA play at first no role, they are however
crucial to provide stability to the chiral soliton: the soliton adds to the FM state energy $E_{\mathrm{FM}}$ a term
\begin{equation}
\Delta E_\mathrm{S} = L_xL_yAq_0\left(8\sqrt{h_y}-2\pi\chi\right).
\end{equation}
The term proportional to $\chi$ comes from the DMI, so that
in absence of DMI the soliton is at most metastable $(\Delta E_\mathrm{S}>0)$, and the solitons of both helicities are degenerated,
having the same energy. The DMI lifts the degeneracy, lowering the energy if $\chi=+1$ and raising it otherwise.
Below the critical field $h_c=\pi^2/16$, the energy of the favored soliton 
becomes negative,  and the proliferation of solitons with the proper helicity ($\chi=+1$)
is energetically favorable. Consequently, they condense forming a CSL~\cite{Dzyal64, Kishine15, Togawa16}.
The properties of isolated solitons play a prominent role in determining the nature of the transition to the CSL phase
\cite{Masaki18,Masaki20}.

To analyze the (meta)stability of a single soliton, let us write the magnetic configuration $\hn$ as
\begin{equation}
\hn = (1-\xi_t^2-\xi_z^2)^{1/2}\hn_0 + \xi_t\hz\times\hn_0 + \xi_z\hz,
\end{equation}
where $\hn_0$ stands for the soliton configuration and $\xi_t$ and $\xi_z$ are two real fields that describe the fluctuations around $\hn_0$.
Expanding the energy in powers of $\xi$ up to second order we get
\begin{equation}
E = E_{\mathrm{FM}}+\Delta E_\mathrm{S} + \int dxdz (\xi_t K_t \xi_t + \xi_z K_z \xi_z)
+ O(\xi^3), \label{eq:taylor}
\end{equation}
where $K_t$ and $K_z$ are the differential operators
\begin{eqnarray}
K_t &=& -\nabla^2 - \frac{1}{2}\varphi_0^{\prime\,2} + q_0^2h_y, \\
K_z &=& -\nabla^2 - \frac{3}{2}\varphi_0^{\prime\,2} + q_0\varphi_0^\prime + q_0^2(h_y-\kappa),\quad
\end{eqnarray}
The terms linear in $\xi$ are absent in Eq.~(\ref{eq:taylor}) since the soliton is a stationary point of the energy.
The soliton is metastable if $K_t$ and $K_z$ are both (semi)positive definite.

\begin{figure}[t!]
\centering
\includegraphics[width=0.23\textwidth]{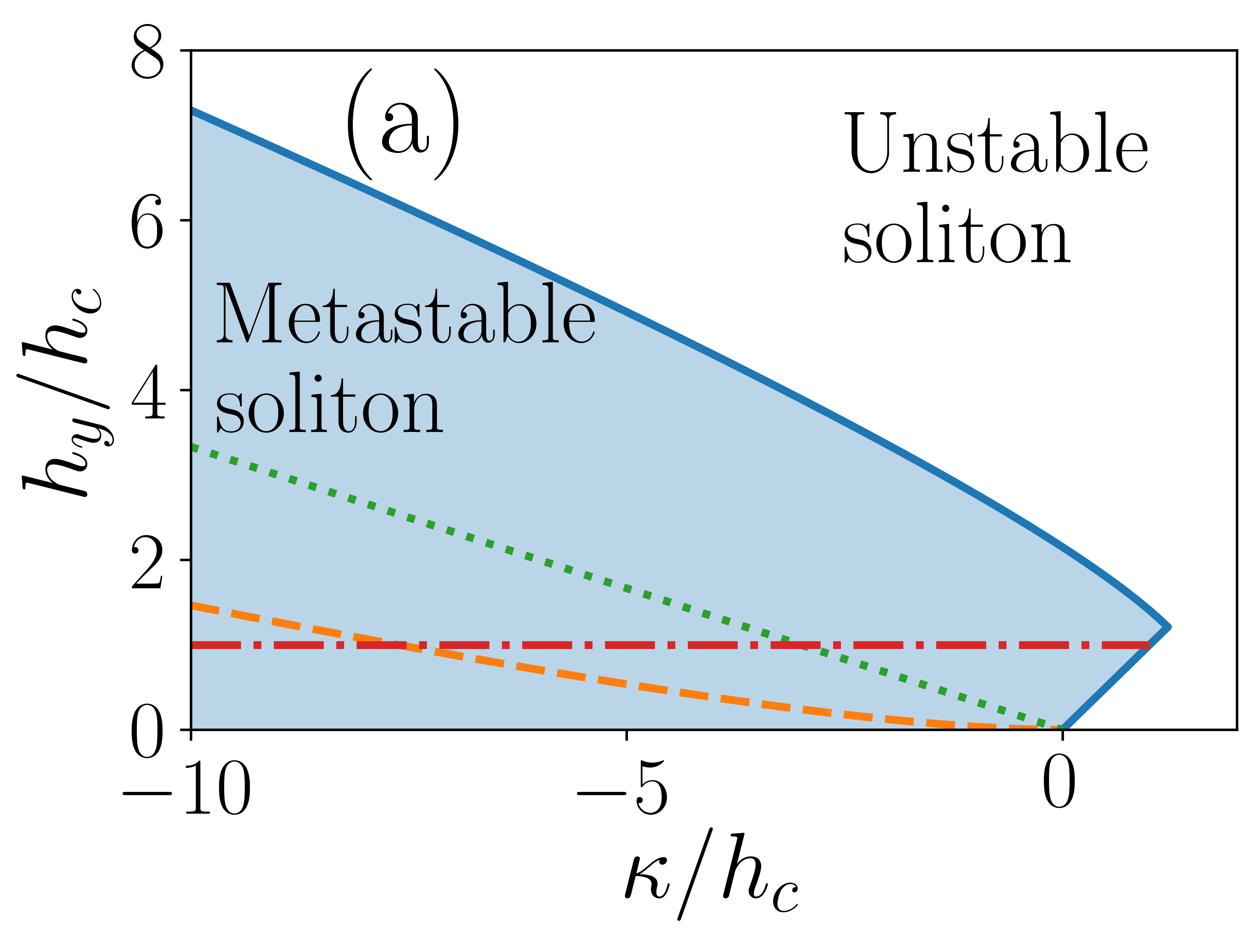}\*
\includegraphics[width=0.23\textwidth]{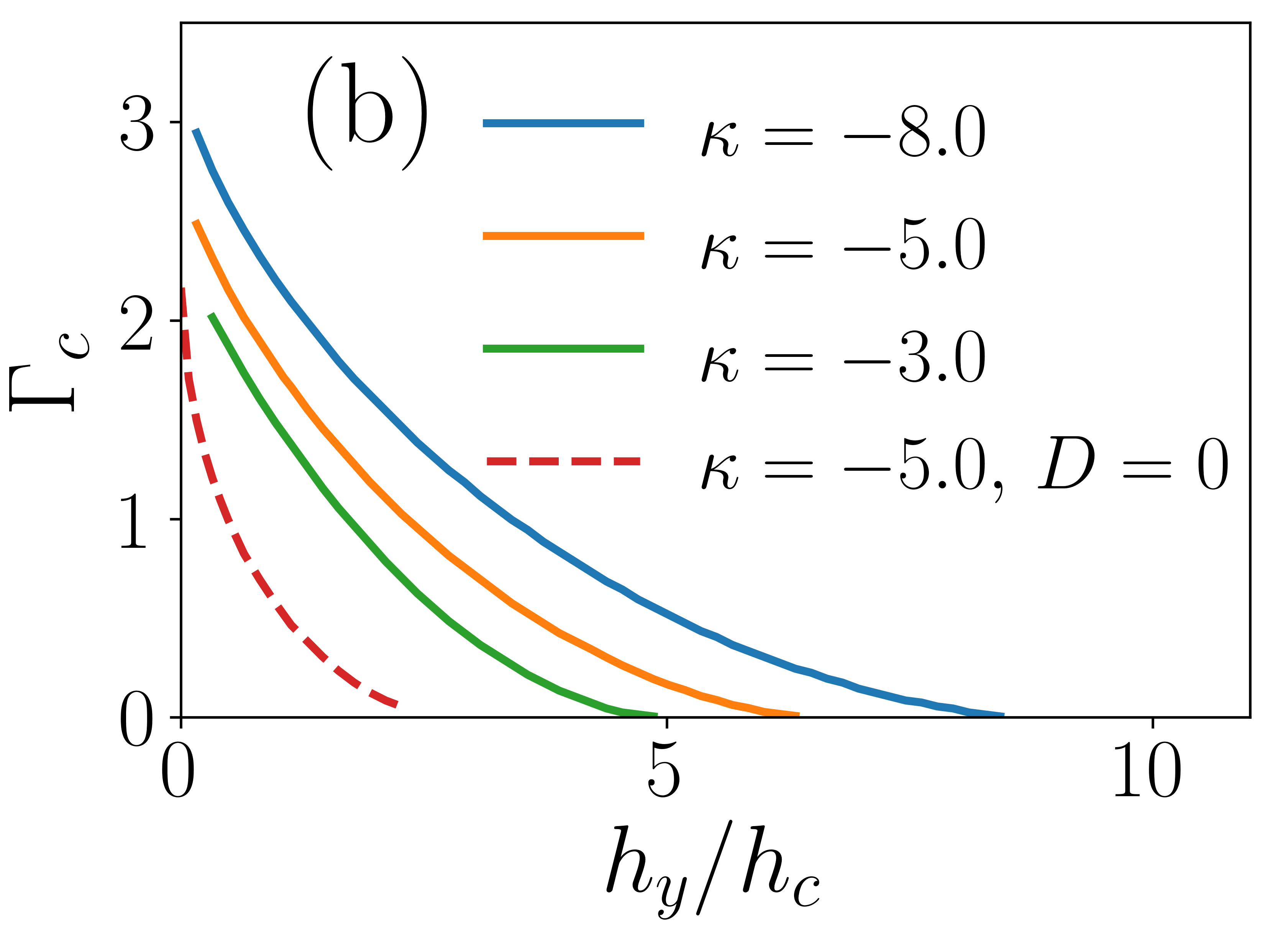}\*
\caption{
  (a) Stability diagram of the chiral soliton for $D>0$, as a function of anisotropy and applied field. The blue continuous line and the orange dashed line correspond
  to the stability limit for $\chi = +1$ and $\chi = -1$, respectively. The green dotted line is the stability limit for $D = 0$ and $\chi = \pm 1$. Below the
  red dash-dotted line the FM state is itself metastable, the ground state being a CSL.
  (b) The critical $\Gamma_\mathrm{c}$ value, proportional to the critical current density, as a function of $h_y/h_c$ for $\chi = +1$ and for several
  values of $\kappa$, as indicated. The red dashed line corresponds to $D=0$.
\label{fig:2}}
\end{figure}

The spectrum of $K_z$ and $K_t$ is studied in detail in the Supplemental Material \cite{suppl}.
It is easy to verify that $K_t$ is always semidefinite positive, so that the soliton stability is determined by the lowest lying eigenvalue of $K_z$.
Without DMI, the solitons of both chiralities are metastable if $h_y<-\kappa/3$. As expected, the DMI enlarges the stability domain of the
$\chi=+1$ soliton and shrinks it if $\chi=-1$.
The stability domains in the $(\kappa/h_c,h_y/h_c)$ plane are represented in Fig.~\ref{fig:2}(a).
The shaded region is the stability domain for $\chi=+1$.
The dashed and dotted lines represent, respectively, the stability boundary for $\chi=-1$ and for both helicities in absence of DMI.
Without anisotropy, the $\chi = +1$ soliton is stable for $h_y\lesssim 2.15 h_c$.
For \CrNbS, which has a large anisotropy ($\kappa\approx -5$), the stability region is much broader: $h_y\lesssim 6.5 h_c$.
Therefore, a metastable soliton can be obtained in a broad region of out-of-plane magnetic fields, $B_y$.

Let us turn to the dynamics.
Contrarily to domain walls, the chiral soliton does not move steadily under the application of a constant out-of-plane magnetic field because the symmetry is not
broken and there are no magnetic domains gaining energy with the field. It is however possible to move the soliton steadily by applying a polarized
electric current, with density $\vec{j}$, which delivers the spin transfer torque~\cite{Zhang04, Manchon19}
\begin{equation}
  \vec{\tau} = -b_j(\vec{j}\cdot\nabla)\hn + \beta b_j \hn\times(\vec{j}\cdot\nabla)\hn,
\end{equation}
with $b_j=P\mu_\mathrm{B}/(|e|M_\mathrm{s})$, where $P$ is the polarization degree, $e$ is the electron charge, and $\mu_B$ is the Bohr magneton.
The first term is the reactive (adiabatic) torque and the second term  the dissipative (non-adiabatic) torque~\cite{Thiaville05}, whose strength is
controlled by the nonadiabaticity coefficient $\beta$. The dynamics obeys the Landau-Lifschitz-Gilbert (LLG) equation
\begin{equation}
\partial_t\hn = \gamma\vec{B}_{\mathrm{eff}}\times\hn + \alpha\hn\times\partial_t\hn +\vec{\tau},
\end{equation}
where $\alpha$ and $\gamma$ are the Gilbert damping parameter and the giromagnetic constant, respectively.

We take the current density $\vec{j} = - j \hz$, and look for a steady solution
that moves rigidly with constant velocity, $v$, along the $\hz$ direction.
The general steady solution is characterized by two functions, $\theta(w)$ and $\varphi(w)$, of the variable $w=q_0(z-vt)$, with BCs: 
$\theta(\pm\infty) = \pi/2$, $\varphi(-\infty)=0$ and $\varphi(+\infty) = \chi 2\pi$.
The LLG equations for steady motion can be cast into the form
\begin{eqnarray}
\theta^{\prime\prime} &=& (\varphi^{\prime\,2}-2\varphi^\prime-h_y\cos\varphi)\cos\theta
+ \kappa\sin\theta\cos\theta \nonumber \\
& & - \Omega\theta^\prime + \Gamma\sin\theta\varphi^\prime,
\label{eq:theta} \\
\varphi^{\prime\prime} &=& h_y\sin\varphi - (\varphi^\prime-2)\cos\theta\theta^\prime
- \Gamma\theta^\prime - \Omega\sin\theta\varphi^\prime \qquad
\label{eq:phi}
\end{eqnarray}
where now the primes stand for derivatives with respect to $w$ and
\begin{equation}
\Omega = \frac{\alpha}{v_0}\Big(v-\frac{\beta}{\alpha}b_jj\Big), \quad
\Gamma = \frac{1}{v_0} \big(v - b_j j\big), \label{eq:OG}
\end{equation}
with $v_0 = 2\gamma Aq_0/M_\mathrm{s}$.
Notice that the spin transfer torque, the Gilbert damping and the
nonadiabaticity coefficient enter the equations of motion only through the constants $\Omega$ and $\Gamma$.

The Boundary Value Problem (BVP) defined by Eqs.~(\ref{eq:theta}) and~(\ref{eq:phi}) 
and the soliton BCs has no solution in general. 
To obtain a solution it is necessary to impose some relation between $\Omega$ and $\Gamma$,
which in its turn determines a relation between the soliton velocity, $v$, and the
applied current intensity, $v_\mathrm{s}$. To see this, let us split the BVP into two pieces,
one for $w\leq 0$ and another one for $w\geq 0$, with the specified soliton BCs for $w\rightarrow\pm\infty$
supplemented with $\theta=\pi/2+\bar{\theta}_0$ and $\varphi=\pi$ at $w=0$.
These two BVP have generically a solution, and have been numerically solved by a 
relaxation method. A solution of the complete BVP, for
$-\infty<w<\infty$, is obtained from the two restricted BVP if the derivatives
$\theta^\prime$ and $\varphi^\prime$ are continuous at $w=0$.
Generically, these two conditions cannot be simultaneously satisfied
by tuning the single degree of freedom at our disposal, $\bar{\theta}_0$. Hence, we have to
tune $\Omega$ and $\Gamma$ to get the complete solution. It turns out that $\varphi^\prime$
is continuous if and only if $\Omega=0$, whatever $\bar{\theta}_0$, which can be tuned to
enforce the continuity of $\theta^\prime$. Therefore, from Eq. (\ref{eq:OG}) we get
\begin{equation}
v = \frac{\beta}{\alpha}b_jj, \label{eq:v}
\end{equation}
and in this case $\Gamma$ becomes proportional to the current intensity: $\Gamma=(\beta/\alpha-1) b_jj/v_0$.
We see that the steady velocity increases linearly with the current density,
with a mobility $m=(\beta/\alpha)b_j$ which is independent of the system parameters $\kappa$ and $h_y$. 
The same behavior occurs for domain walls \cite{Thiaville05}, and thus this seems to be a universal feature of
the response of one dimensional magnetic solitons to polarized currents.
Eq.~(\ref{eq:v}) implies that $v=0$ if $\beta=0$, so that the steady solution is indeed static if there
is no dissipative torque. In that case $\Gamma=-b_jj/v_0$ and the soliton reaches a
different equilibrium state after applying the current. Finally, the case $\beta=\alpha$ is special,
since $\Omega=0$ and $\Gamma=0$, so that Eqs. (\ref{eq:theta}) and (\ref{eq:phi}) are independent of the applied current.
Hence, the soliton is rigidly dragged by the current, with velocity $v=b_jj$, without changing its static shape.

By increasing the current, $\bar{\theta}_0$ increases from its static value $\bar{\theta}_0=0$.
At sufficiently large $\Gamma$ a second, \textit{unstable}, solution of the BVP, with larger $\bar{\theta}_0$, appears.
At a certain $\Gamma = \Gamma_c$, which depends on the system parameters,
the stable and unstable branches meet and the steady solution becomes unstable~\cite{suppl}.
Thus, no steady moving soliton exists above this critical current.
If $\beta=\alpha$ Eqs. (\ref{eq:theta}) and (\ref{eq:phi}) are independent of the current, and thus there is no critical current.
Again, a similar scenario is observed for moving domain walls~\cite{Thiaville05}.
The critical current decreases with $h_y$ and increases with $\kappa$, as shown in Fig.~\ref{fig:2}(b).
This agrees with the fact that the field tends to destabilize the soliton while DMI and UMA tend to stabilize it.
It is worthwile to stress that the mobility is independent of $\chi$ if the solitons of both helicities are metastable.
However, for given current the soliton profiles depend on $\chi$, and, as expected, the critical current is much smaller
for $\chi=-1$~\cite{suppl}.

\begin{figure}[t!]
\includegraphics[width=0.224\textwidth]{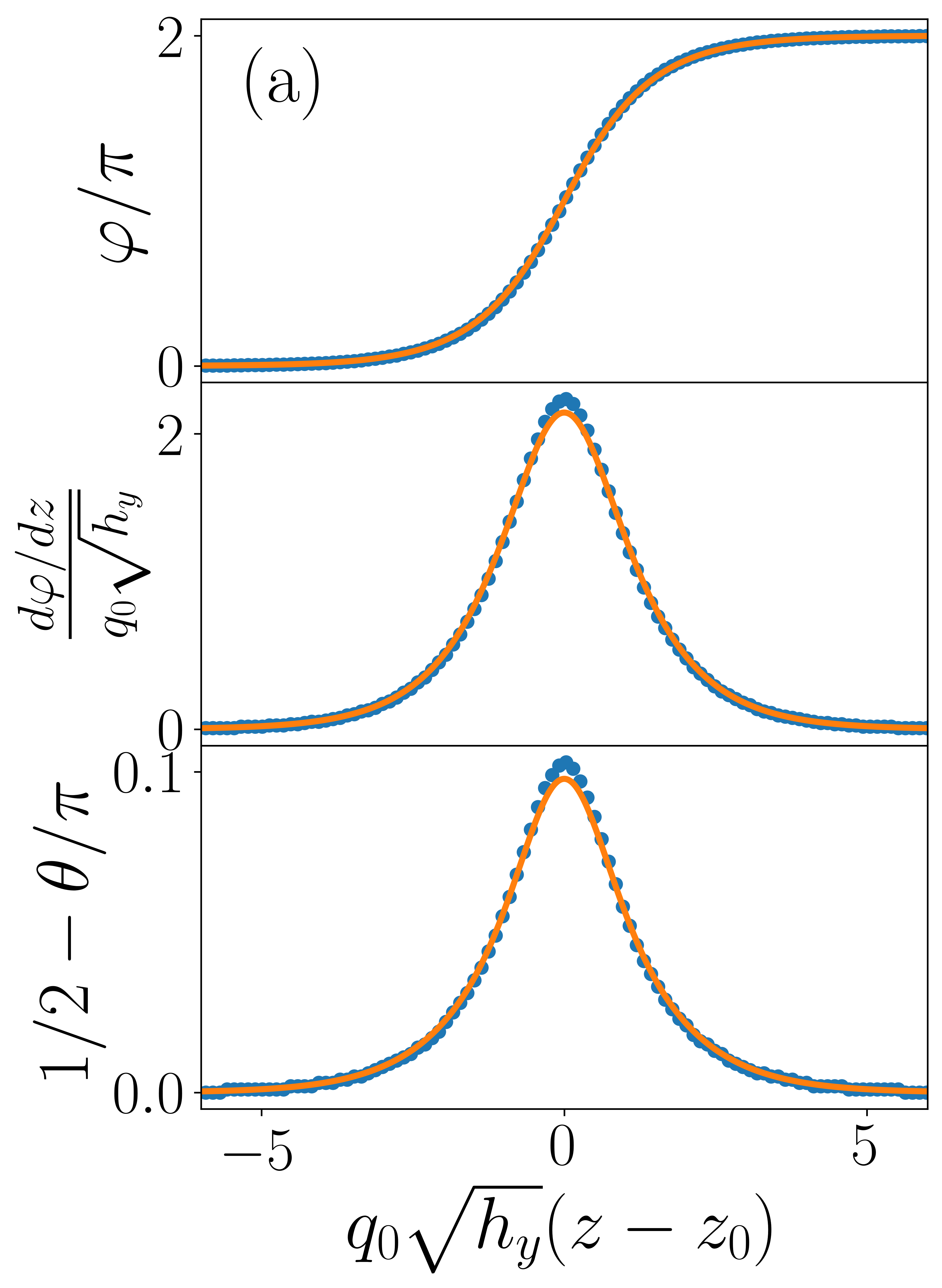}\*
\includegraphics[width=0.236\textwidth]{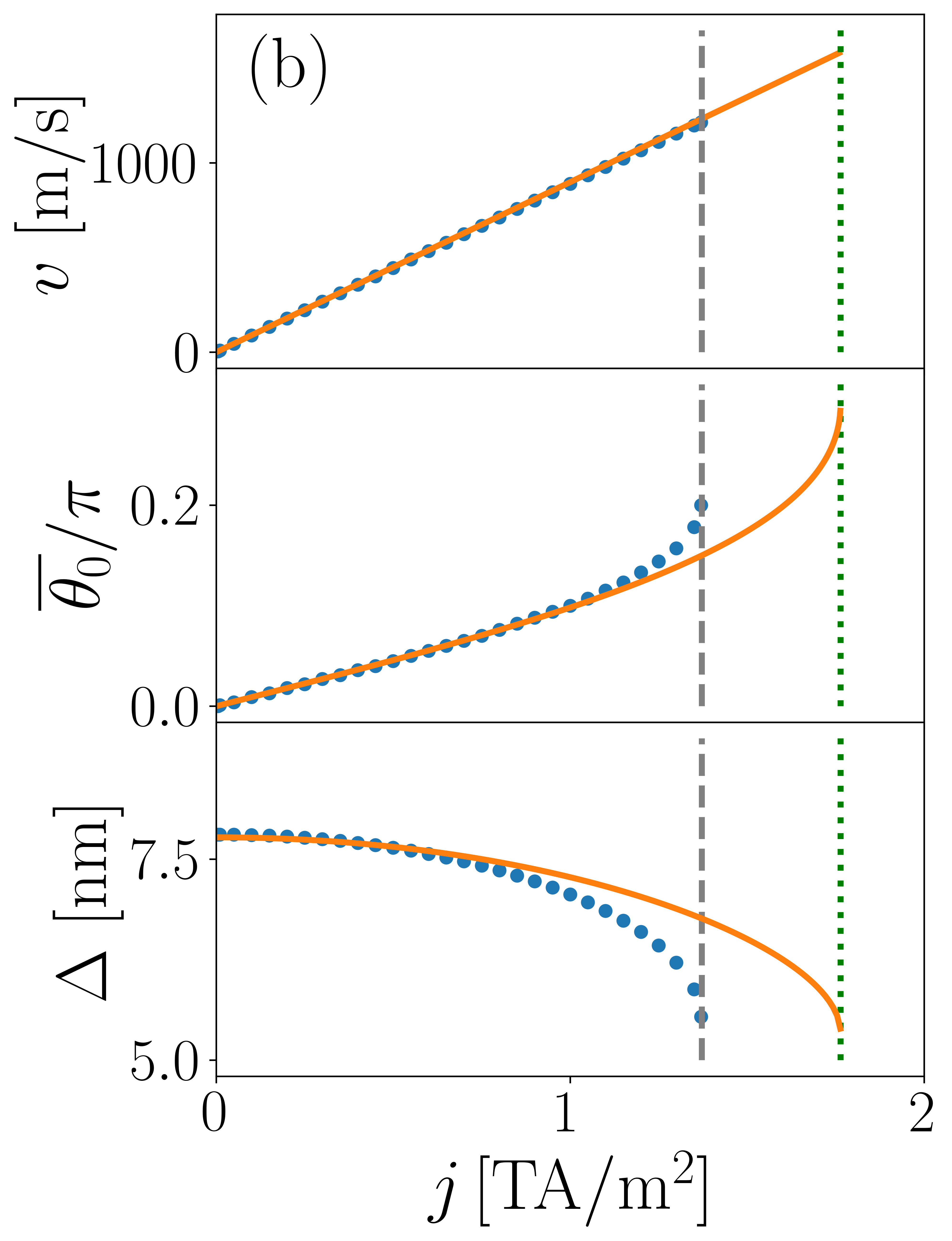}\*
\caption{Steady motion of the chiral soliton.
(a) Steady profiles for $\kappa=-5.17$, $h_y=0.807$, and $\Gamma=0.89$ ($j = 1\,\mathrm{TA/m^2}$).
Circles correspond to numerical simulations and lines to the BVP.
(b) Velocity and soliton parameters as a function of the applied current density $j$. The steady velocity increases linearly
with the current, with mobility $m = (\beta/\alpha) b_j$, as indicated by the continuous line (top panel). Middle and bottom panels:
$\bar{\theta}_0$, (tilt of the magnetization in the $z$ direction) increases with $j$, whilst the soliton width $\Delta$ decreases.
Both quantities show a considerable change when the critical current $j_c = 1.372\,\mathrm{TA/m^2}$,
indicated by the vertical dotted line, is approached.
Continuous lines correspond to the solution of the corresponding BVP. Vertical dashed and dotted lines correspond to the critical
values $j_c$ obtained with numerical simulations and with the BVP, respectively.
\label{fig:3}}
\end{figure}

Two important questions are not addressable by the BVP: 1) the fate of the soliton for $j>j_c$, and 2) whether the steady moving regime
is reached by applying a current to a static soliton.
To answer these questions, we performed numerical simulations of the LLG equation using the MuMax3 code \cite{MuMax3, Exl15, Leliaert18},
in which we have implemented the monoaxial DMI~\cite{suppl}.
We used the following parameters, appropriate to \CrNbS:
$A=1.42\,\mathrm{pJ/m}$, $D = 369\,\mathrm{\mu J/m^2}$,
$K=-124\,\mathrm{kJ/m^3}$,
$M_\mathrm{S}=129\,\mathrm{kA/m}$, \mbox{$\alpha=0.01$}, \mbox{$\beta=0.02$}, and \mbox{$P=1$}.
In addition, we set $B_y=300\,\mathrm{mT}$ which is larger than the stability limit of the CSL $B_{y,\mathrm{c}} = 230\,\mathrm{mT}$.
These values correspond to \mbox{$q_0=0.13$ nm$^{-1}$}, \mbox{$\kappa=-5.17$} and \mbox{$h_y=0.807$}.
The numerical solution of the BVP for this set of parameters gives $\Gamma_c = 1.5735$ (see Fig.~\ref{fig:2}).
As a test of the code, we have obtained that, in absence of applied magnetic field, the system relaxes to a helical state with wave number $q_0$,
and that a metastable chiral soliton can be retained for a broad $h_y$ range.

The shape of the steady moving solution for $\chi=+1$ is displayed in Fig.~\ref{fig:3}(a) for $j = 1\,\mathrm{TA/m^2}$ ($\Gamma=0.89$). 
Continuous lines correspond to the solution of the BVP and circles to the steady profile found by numerical simulations of the LLG equation,
showing good agreement between them.
The bottom panel in Fig.~\ref{fig:3}(a) shows that the magnetization in the soliton is tilted towards $\hz$, the direction of the velocity.
Let $z_0$ be the center of the soliton, given by the maximum of $\varphi^\prime$, where now the prime stands again for derivative respect to $z$.
The tilt angle is the deviation of the polar angle from $\pi/2$ in its center, $\bar{\theta}_0 = \pi/2 - \theta(z_0)$.
The soliton width $\Delta$ can be defined in terms of $\varphi'$ as 
$\Delta^2 = \int (z-z_0)^2 \varphi'(z)^2dz/\int \varphi'(z)^2dz$, 
The values of $\Delta$ and $\bar{\theta}_0$ depend on the applied current density and on the system parameters.
Figure \ref{fig:3}(b) displays the steady velocity, $v$, and $\bar{\theta}_0$ and $\Delta$ as a function of $j$.

\begin{figure}[t!]
\includegraphics[width=0.234\textwidth]{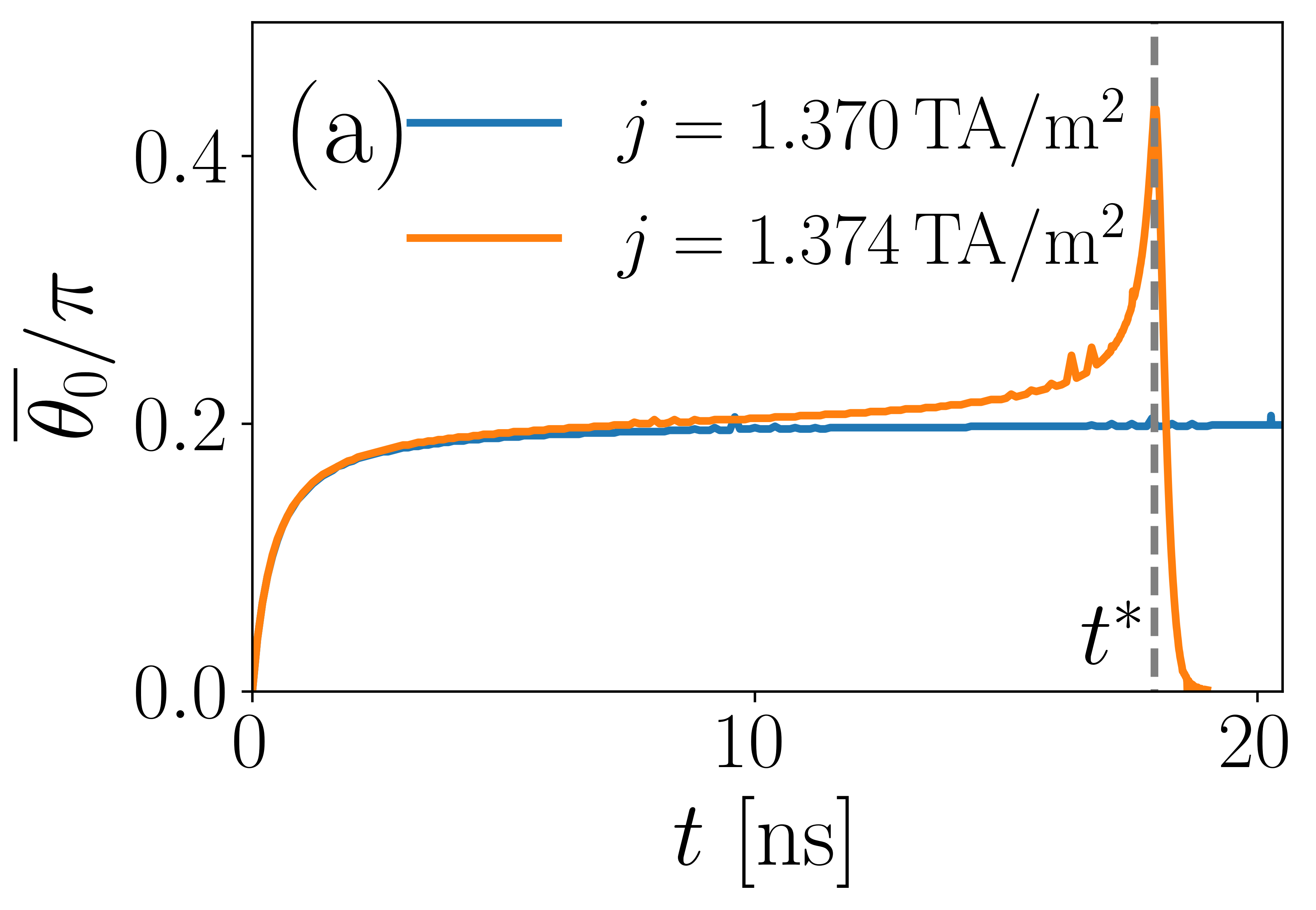}\*
\includegraphics[width=0.226\textwidth]{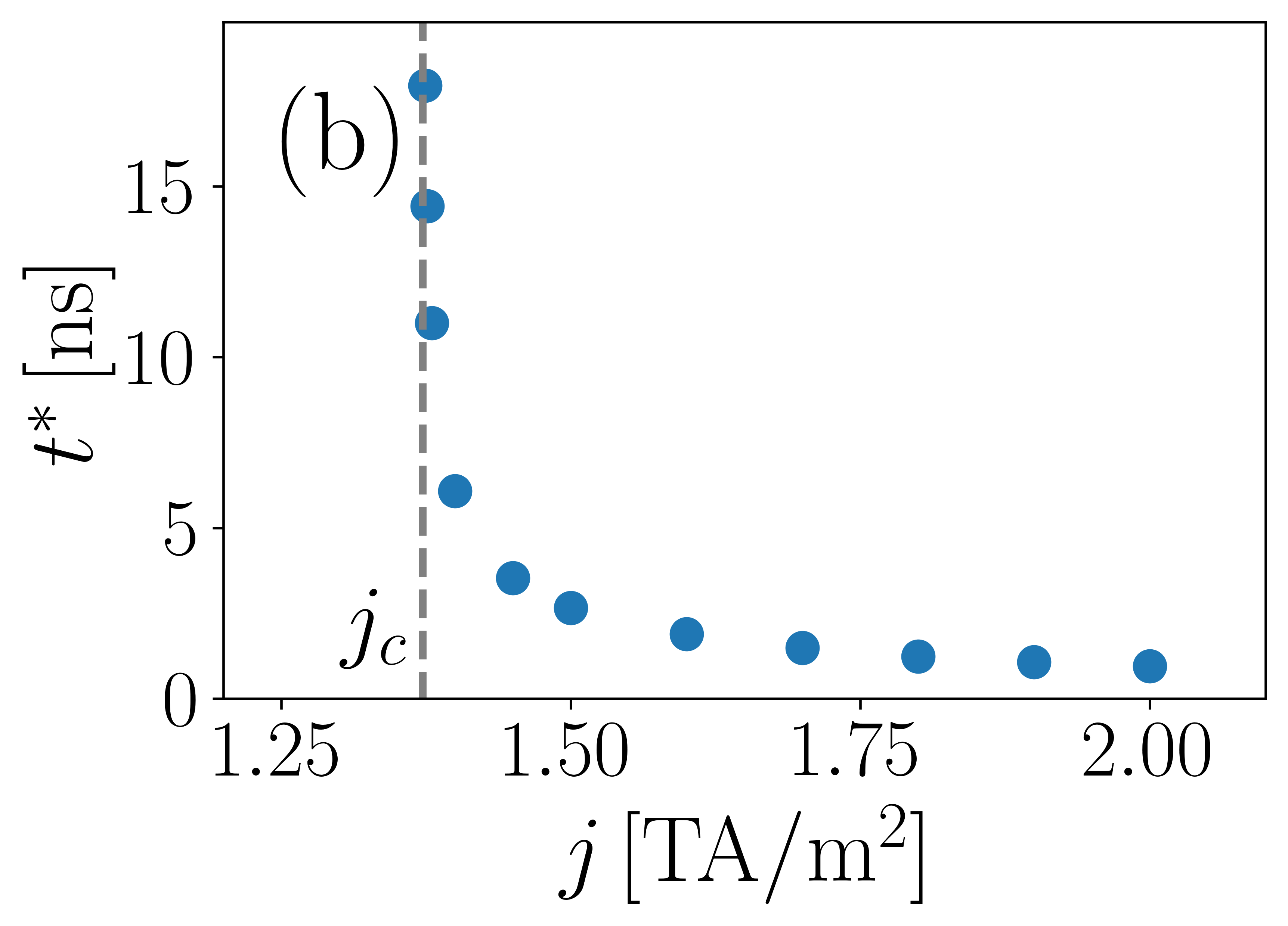}\*
\caption{Instability of the chiral soliton under an applied current density.
  (a) Evolution with time of the tilt angle $\bar{\theta}_0$ around the critical current $j_c$.
  The vertical line indicates the value of $t^*$ for $j= 1.374\,\mathrm{TA/m^2}$, beyond which the magnetization in the center of the soliton abruptly goes to the $y$ direction.
  (b) Dependence on the current density of the instability time $t^*(j)$, showing how it seems to diverge when approaching $j_c = 1.372\,\mathrm{TA/m^2}$ from above.
\label{fig:4}}
\end{figure}

Numerical simulations show that the system, starting from the metastable static soliton, reaches the steady motion state if the current is below
the critical current, $j_\mathrm{c} = 1.372\,\mathrm{TA/m^2}$, which corresponds to $\Gamma = 1.224$, slightly smaller than the value
of $\Gamma_c$ predicted with the BVP. Currents higher than $j_\mathrm{c}$ destroy the soliton and drives the system to the FM state.
Fig.~\ref{fig:4} displays results of numerical simulations that clarify the fate of the soliton upon application of a supercritical current.
Fig.~\ref{fig:4}(a) presents the temporal evolution of $\bar{\theta}_0$ for $j=1.370\,\mathrm{TA/m^2}$, for which a steady soliton motion is
reached, and for $j=1.374\,\mathrm{TA/m^2}$, where no steady solitonic state is attained at long times.
The dotted vertical line indicates the time $t^* = 17.95$ ns when the soliton is destroyed, which is anticipated by the sudden increase
of $\bar{\theta}_0$. The dependence of $t^*$ on the value of the supercritical current density is presented in Fig~\ref{fig:4}(b),
showing how it tends to diverge when reaching $j_c$ from above. Beyond $t^*$, the system goes to a FM state with the magnetization completely oriented along
the direction of the external field~\cite{suppl}.

The critical current resembles the one appearing in domain walls \cite{Thiaville05} and is tantamount
to the Walker breakdown field \cite{Walker56,Schryer74,Krizakova19}. However, currents beyond the Walker breakdown do not destroy the domain wall, but induce
a non-steady precesional motion. This is a major difference between the chiral soliton and domain wall steady motion. The destruction of
the soliton by supercritical currents can be a very useful tool to manipulate information in potential spintronic devices that use the
presence or absence of solitons as bits.

In summary, we have shown that single chiral solitons can be metastably retained in monoaxial helimagnets and that they can be controlled by
applying a polarized electric current.
The metastability of the soliton is guaranteed by the DMI interaction and the UMA.
The steady velocity is proportional to the current density, with a
mobility given by the ratio between the nonadiabaticity and the Gilbert damping coefficients.
Notably, the soliton is destabilized when a critical current density value is reached.
This controlled motion of chiral solitons opens a new route to the development of spintronic devices based in
topological structures.

\begin{acknowledgments}
SB acknowledges interesting discussions with J. Curiale.
Grants No PGC-2018-099024-B-I00-ChiMag from the Ministry of Science and Innovation
of Spain, i-COOPB20524 from CSIC, DGA-M4 from the Diputación General de Aragón, and PICT
2017-0906 from the Argentinian Agencia Nacional de Promoci\'on Cient\'{\i}fica y Tecnol\'ogica, are acknowledged.
This work was also supported by the MEXT program for promoting the enhancement of research 
universities, by the JSPS Core-to-Core Program, A. (Advanced Research Networks), by the 
Chirality Research Center (Crescent) in Hiroshima University, and by JSPS and RFBR under the 
Japan - Russia Research Cooperative Program. 
\end{acknowledgments}

\bibliography{references}

\end{document}



\title{Supplemental material to
  ``Dynamics of chiral solitons driven by polarized currents in monoaxial helimagnets''}


\author{Victor Laliena}
\email[]{laliena@unizar.es}
\affiliation{Aragon Material Science Institut 
  (CSIC -- University of Zaragoza) and Condensed Physics Matter Department, University of Zaragoza \\
  C/ Pedro Cerbuna 12, 50009 Zaragoza, Spain}
\author{Sebastian Bustingorry}
\email[]{sbusting@gmail.com}
\affiliation{
Instituto de Nanociencia y Nanotecnolog\'{\i}a, CNEA-CONICET, Centro At\'omico Bariloche, (R8402AGP), S. C. de Bariloche, R\'{\i}o Negro, Argentina}
\author{Javier Campo}
\email[]{javier.campo@csic.es}
\affiliation{Aragon Material Science Institut 
  (CSIC -- University of Zaragoza) and Condensed Physics Matter Department, University of Zaragoza \\
  C/ Pedro Cerbuna 12, 50009 Zaragoza, Spain}


\date{July 15, 2020}

\maketitle


The supplemental material contais details about computations of the chiral soliton stability domain,
of the BVP that provides the steady solution, of the implementation of monoaxial DMI in the
MuMax program and numerical simulations, and of the dynamics of the unfavored $\chi = -1$ chiral soliton.

\section{\noindent S1. Chiral soliton stability}

The chiral soliton is metastable if the differential operators $K_t$ and $K_z$
defined by Eqs. (8) and (9) in the main text are (semi)positive definite. Let us write them here
again for the reader convenience:
\begin{eqnarray}
K_t &=& -\nabla^2 - \frac{1}{2}\varphi_0^{\prime\,2} + q_0^2h_y, \\
K_z &=& -\nabla^2 - \frac{3}{2}\varphi_0^{\prime\,2} 
+ q_0\varphi_0^\prime + q_0^2(h_y-\kappa),\quad
\end{eqnarray}
where the prime stands for derivative respect to $z$.
By Fourier transform in $x$ and $y$, and changing the variable to $w=\chi\sqrt{h_y}q_0z$,
the eigenvalues of $K_t$ and $K_z$, denoted by $\mu_t$ and $\mu_z$ respectively, 
can be written as
\begin{eqnarray}
\mu_t &=& q_0^2h_y(\lambda_t+1) + k_x^2 + k_y^2\\
\mu_z &=& q_0^2h_y(\lambda_z+1) - q_0^2\kappa + k_x^2 + k_y^2 \label{eq:muz}
\end{eqnarray}
where $k_x$ and $k_y$ are the components of the Fourier wave vector, and
$\lambda_t$ and $\lambda_z$ are the eigenvalues of
\begin{eqnarray}
\tilde{K}_t &=& -\frac{d^2}{dw^2} - \frac{2}{\cosh^2w}, \\
\tilde{K}_z &=& -\frac{d^2}{dw^2} - \frac{6}{\cosh^2w} 
+ \frac{\chi}{\sqrt{h_y}}\frac{4}{\cosh{w}},
\end{eqnarray}
respectively. From now on the prime stands for the derivative with respect to $w$.

Let us consider first $\tilde{K}_t$, which is the Schr\"odinger operator with a
P\"oschl-Teller potential. Its lowest lying eigenvalue is \mbox{$\lambda_t=-1$}
\cite{Poschl33}. Hence, $\mu_t\geq 0$ and $K_t$ is always semidefinite positive.
This result can be quickly obtained by noticing that $\varphi_0^\prime$
is an eigenstate of $\tilde{K}_t$ with zero eigenvalue. This can be checked
either by direct application of $\tilde{K}_t$ to $\varphi^\prime$, and it is due
to the fact that $\varphi^\prime$ is the generator of infinitesimal translations
of the soliton along $z$:
\begin{equation}
\hn_0(w+\delta w) = \hn_0(w) + \delta w \varphi_0^\prime \hz\times\hn_0(w),
\end{equation}
and $\hn_0(w+\delta w)$ and $\hn_0(w)$ have the same energy.
Given that $\varphi_0^\prime$ has no nodes (it does not vanish at any point),
Sturm theorem \cite{Dunford64} guarantees that it corresponds to the lowest
lying eigenvalue.

\begin{figure}[t!]
\centering
\includegraphics[width=0.6\textwidth]{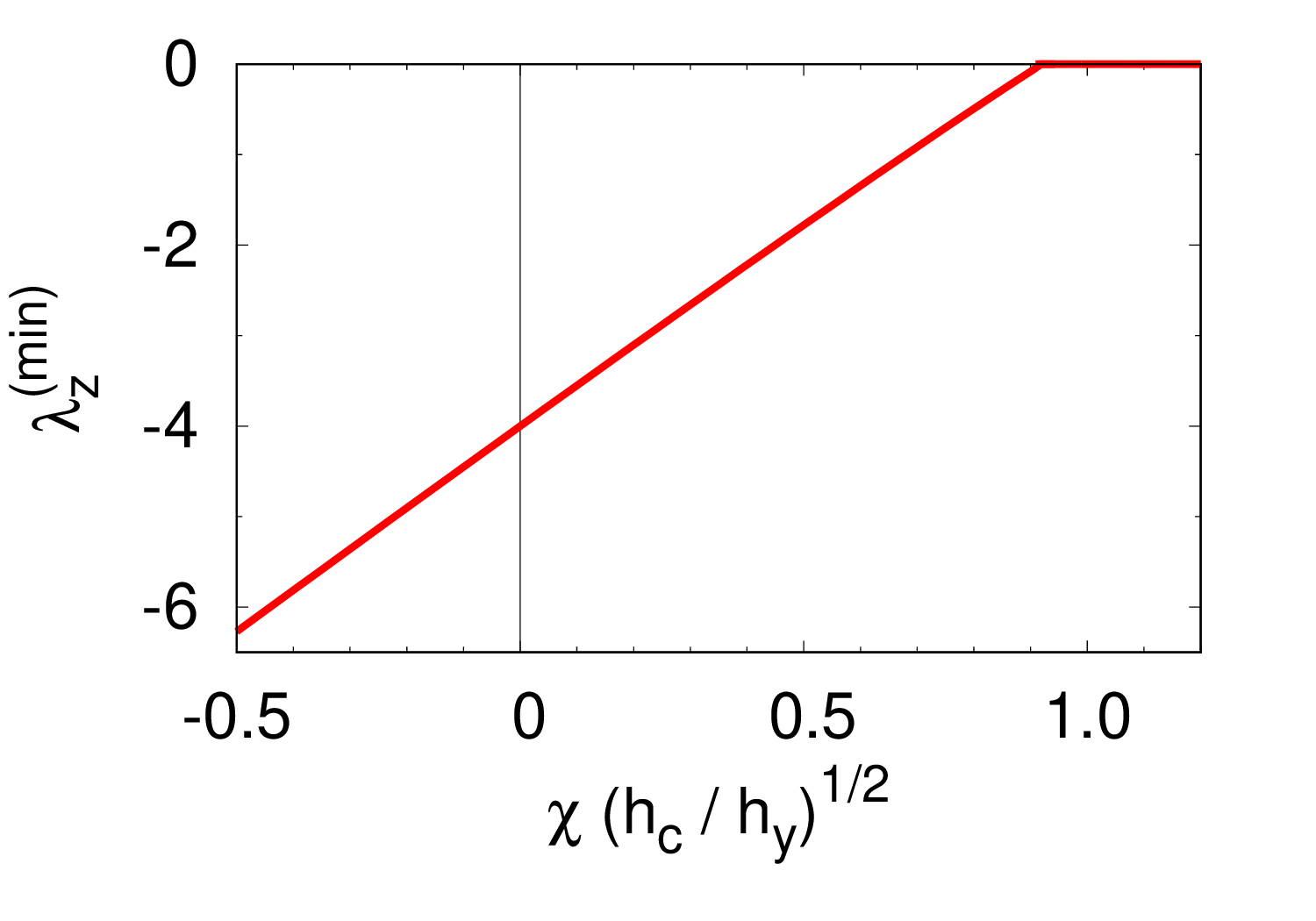}
\caption{Lowest lying eigenvalue of $\tilde{K}_z$.
\label{fig:1}}
\end{figure}

Hence, the stability of the soliton is solely determined by the lowest lying
eigenvalue of $\tilde{K}_z$, denoted by $\lambda_z^{\mathrm{(min)}}$.
For given $\chi$ and $h_y$, we have $\mu_z\geq 0$
if and only if $\kappa\leq\kappa_c$, with
\begin{equation}
\kappa_c = h_y\big(\lambda_z^{\mathrm{(min)}}+1\big). \label{eq:kappac}
\end{equation}
Notice that the spectrum of $\tilde{K}_z$ depends only on $\chi/\sqrt{h_y}$,
or, equivalently, on $\chi\sqrt{h_c/h_y}$, where $h_c=\pi^2/16$ is the
critical field for soliton proliferation.
If we set $\chi=0$ in $\tilde{K}_z$, it becomes a P\"oschl-Teller potential
whose lowest lying eigenvalue is $\lambda_z^{\mathrm{(min)}}=-4$. Hence,
in absence of DMI, the chiral soliton is stable for $h_y<-\kappa/3$, and, 
in a simple ferromagnet, without DMI and anisotropy, it is always unstable,
since $h_y\geq 0$ by definition.
The lowest lying eigenvalue of $\tilde{K}_z$ as
a function of $\chi\sqrt{h_c/h_y}$ is displayed in Fig.~\ref{fig:1}.
The soliton (metas)tability boundary as a function of $\chi$, $\kappa$, and $h_y$
can be determined from Eq.~(\ref{eq:kappac}) and Fig.~\ref{fig:1}. 

\section{\noindent S2. Solution of the Boundary Value Problem}

The soliton steady motion driven by a polarized torque is determined
by the Boundary Value Problem (BVP) set by equations (12) and (13) of the main text,
which we reproduce below for the reader convenience,
\begin{eqnarray}
\theta^{\prime\prime} &=& (\varphi^{\prime\,2}-2\varphi^\prime-h_y\cos\varphi)\cos\theta
+ \kappa\sin\theta\cos\theta - \Omega\theta^\prime + \Gamma\sin\theta\varphi^\prime,
\label{eq:theta} \\
\varphi^{\prime\prime} &=& h_y\sin\varphi - (\varphi^\prime-2)\cos\theta\theta^\prime
- \Gamma\theta^\prime - \Omega\sin\theta\varphi^\prime,
\label{eq:phi}
\end{eqnarray}
and the boundary conditions (BCs)
\begin{equation}
\theta(\pm\infty)=\pi/2,\quad \varphi(-\infty)=0, \quad \varphi(+\infty)=\chi 2\pi,
\end{equation}
where $\theta$ and $\varphi$ are functions of the variable $w = q_0(z-vt)$
and the prime stands for the derivative with respect to $w$.

\begin{figure}[t!]
\centering
\includegraphics[width=0.24\textwidth]{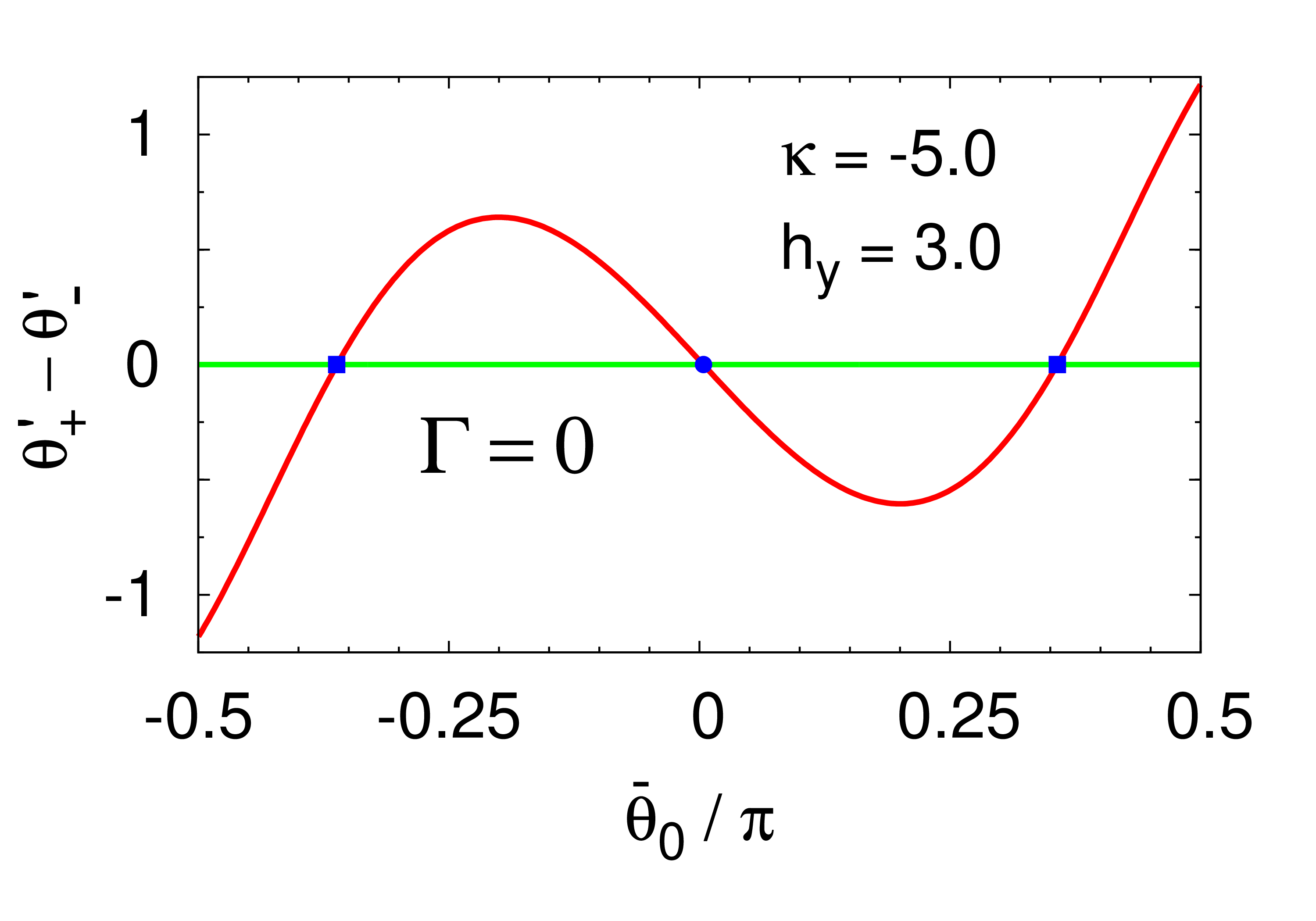}
\includegraphics[width=0.24\textwidth]{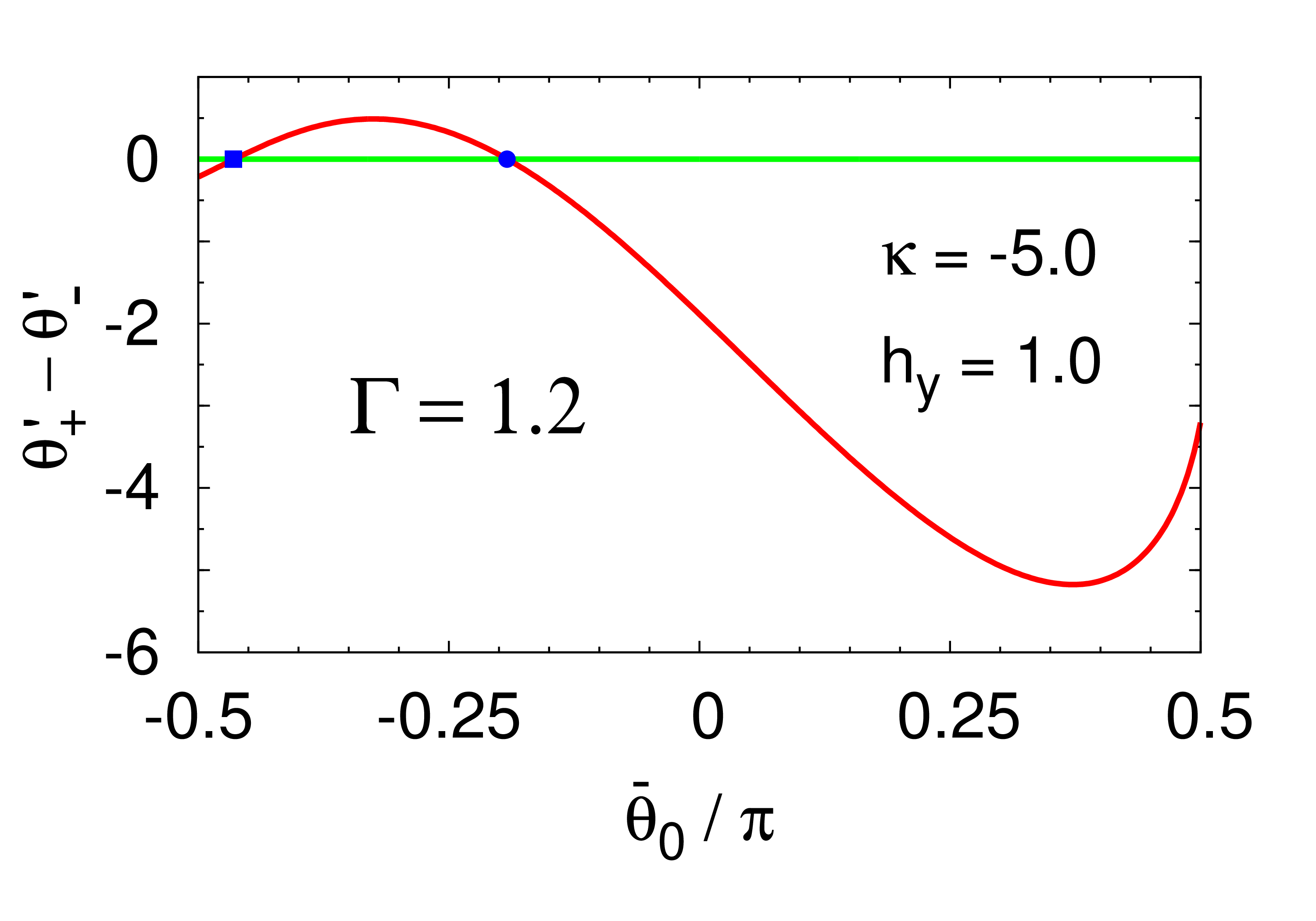}
\includegraphics[width=0.24\textwidth]{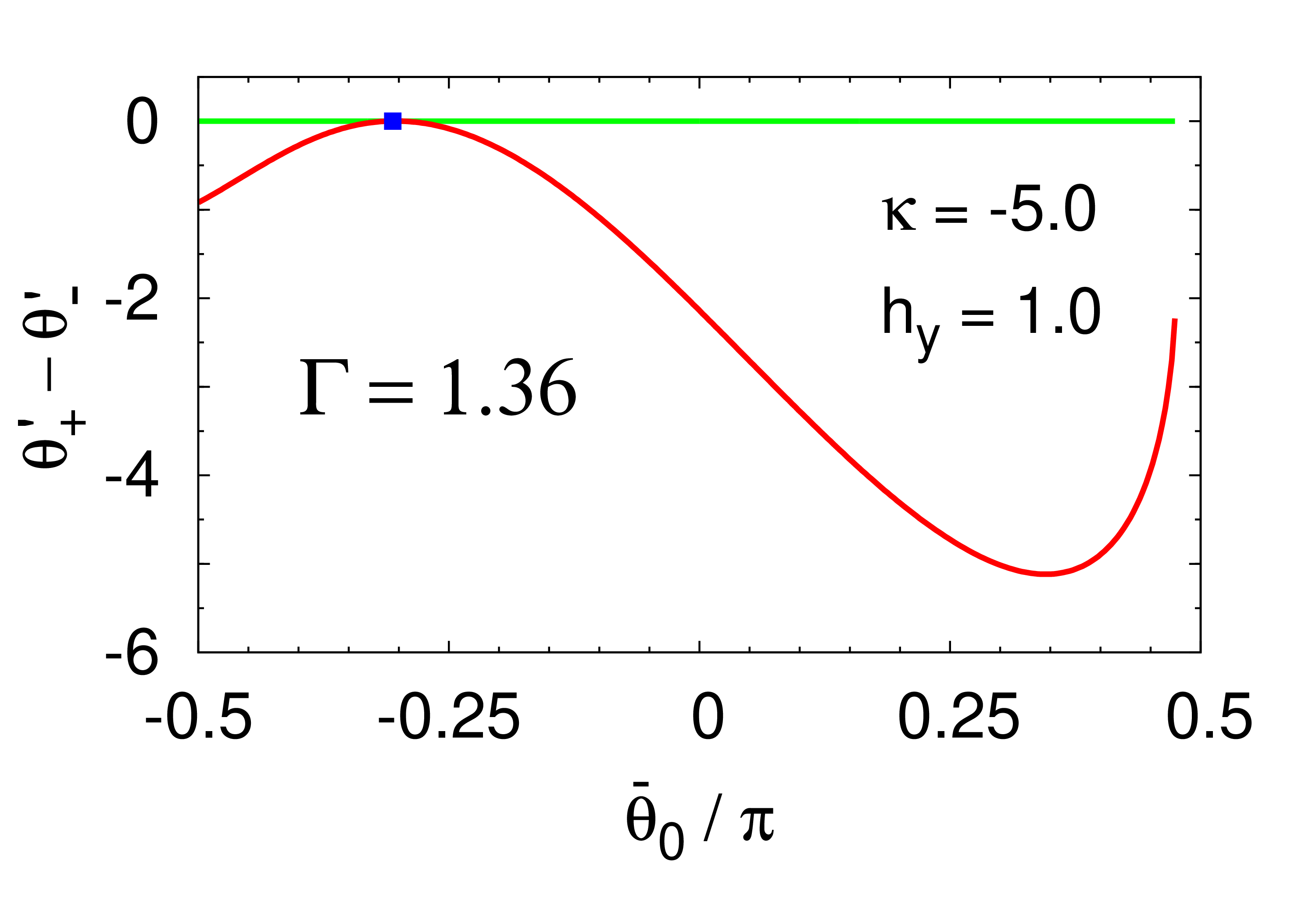}
\includegraphics[width=0.24\textwidth]{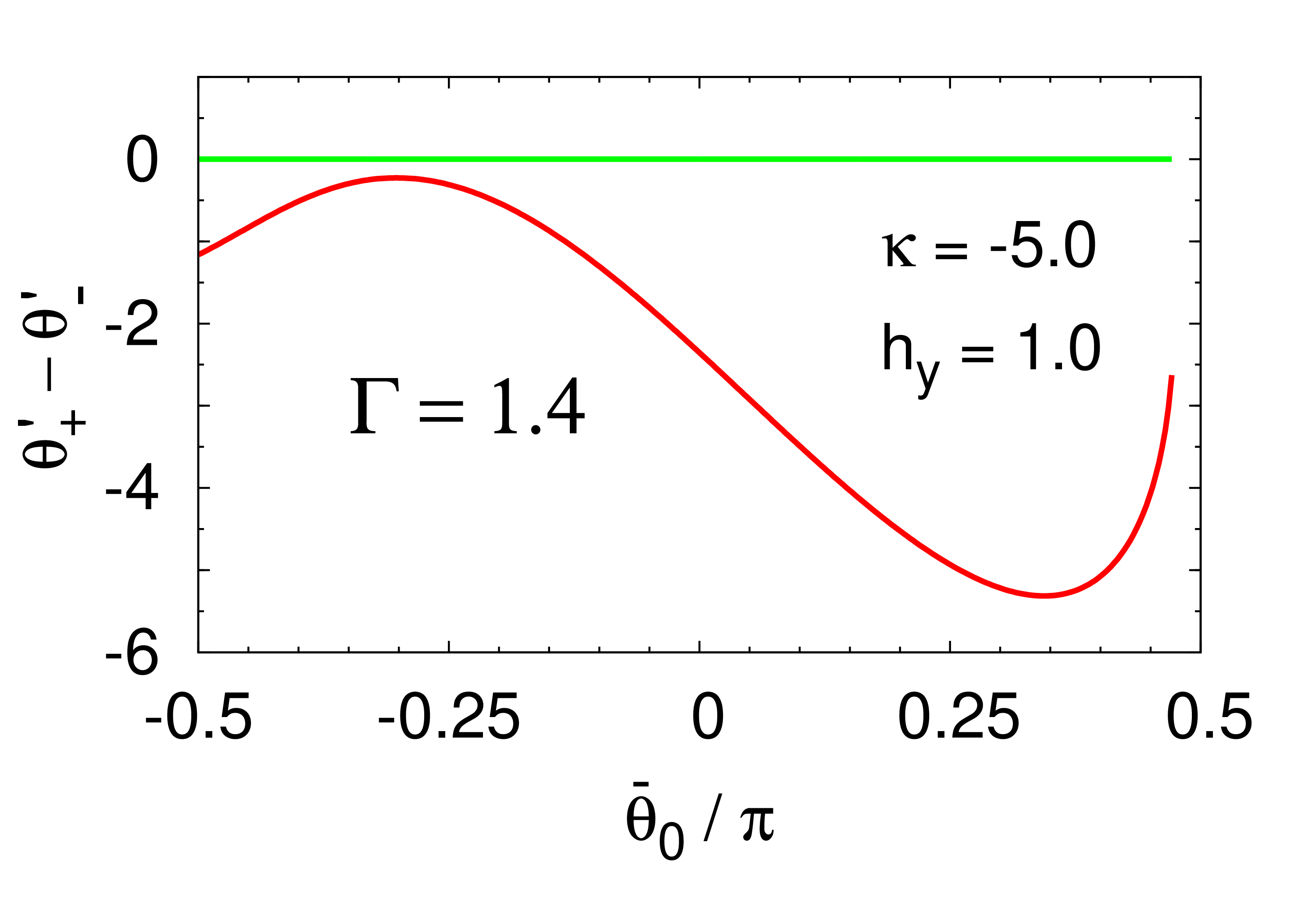} \\
\includegraphics[width=0.24\textwidth]{scanB0.png}
\includegraphics[width=0.24\textwidth]{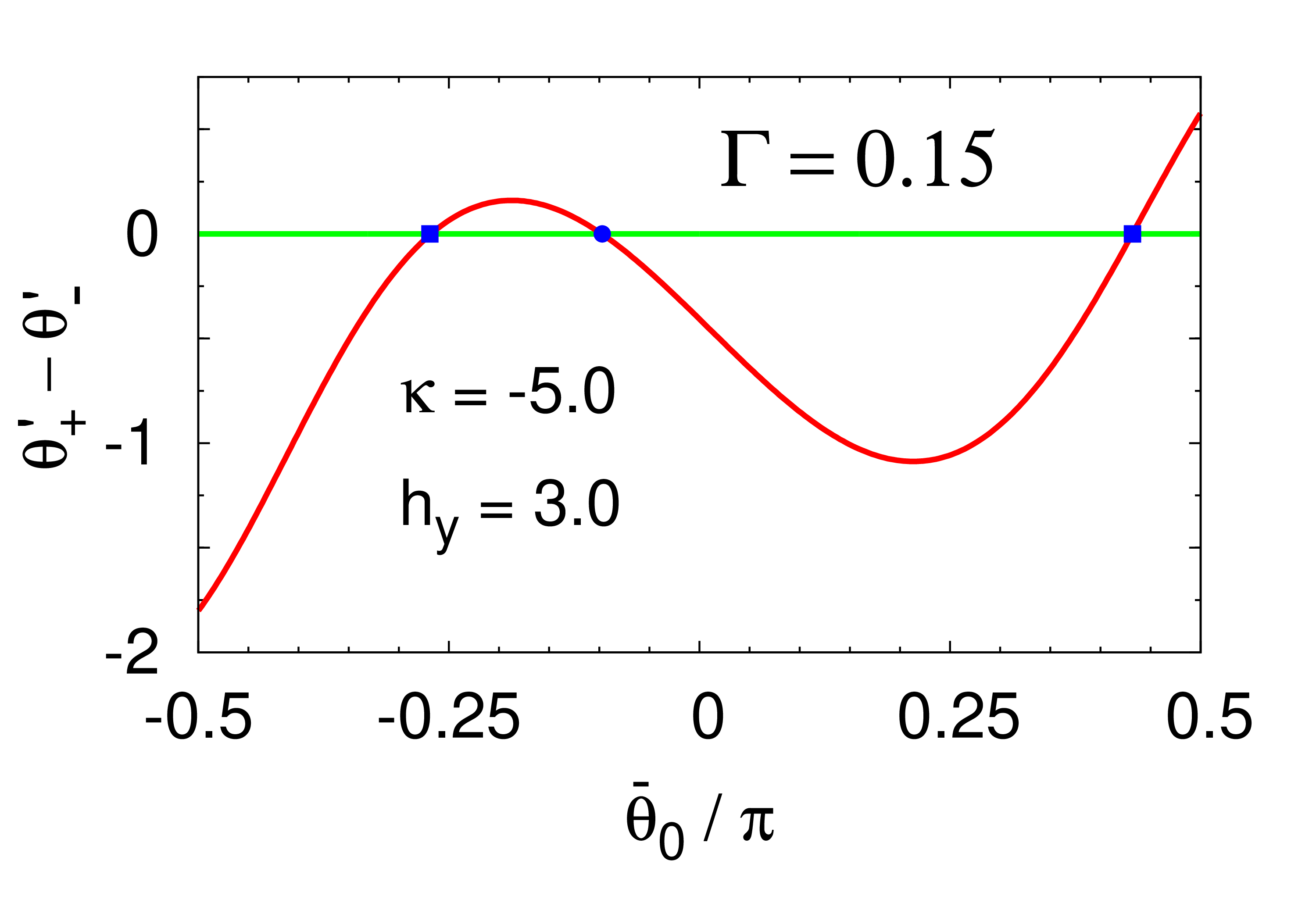}
\includegraphics[width=0.24\textwidth]{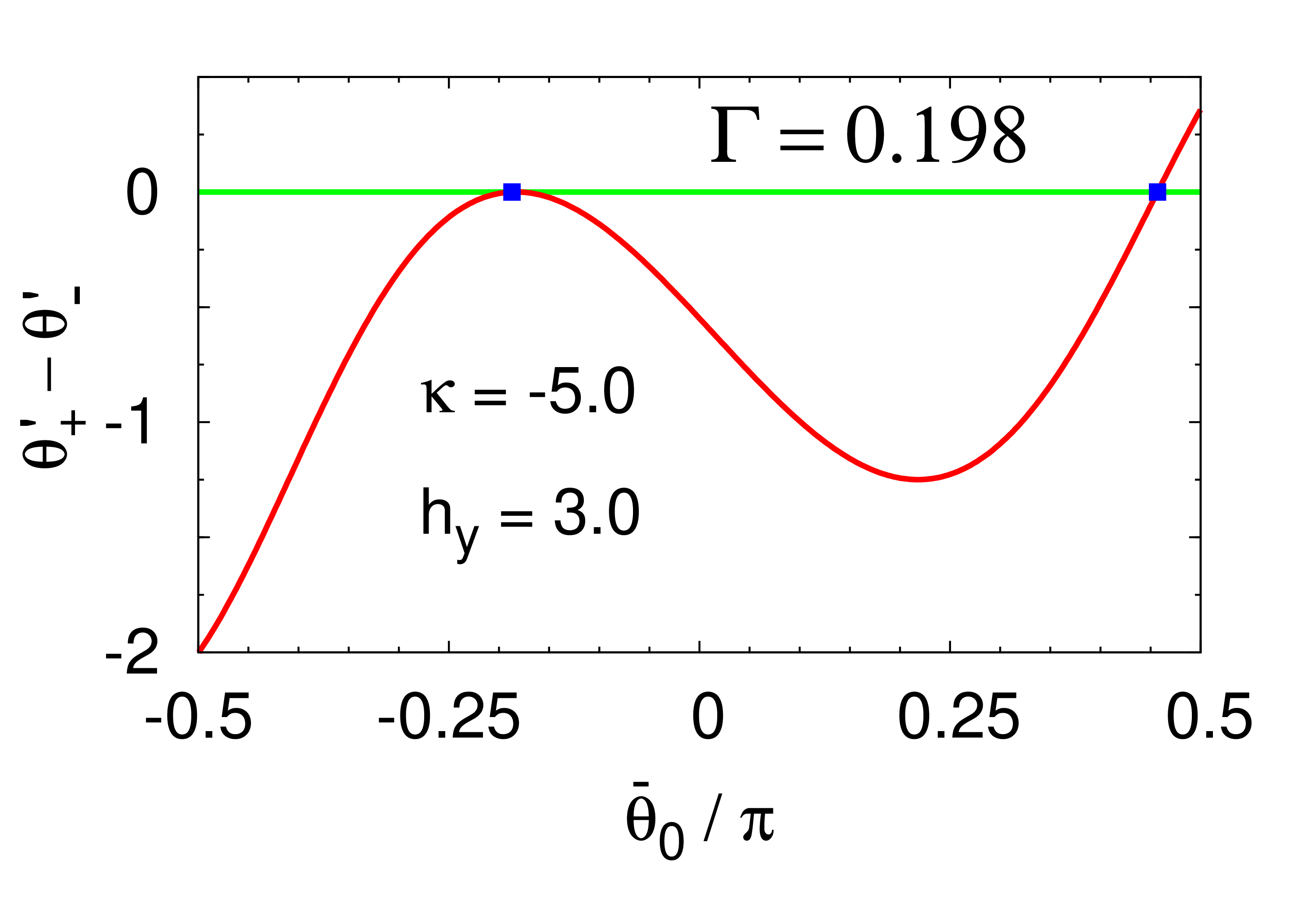}
\includegraphics[width=0.24\textwidth]{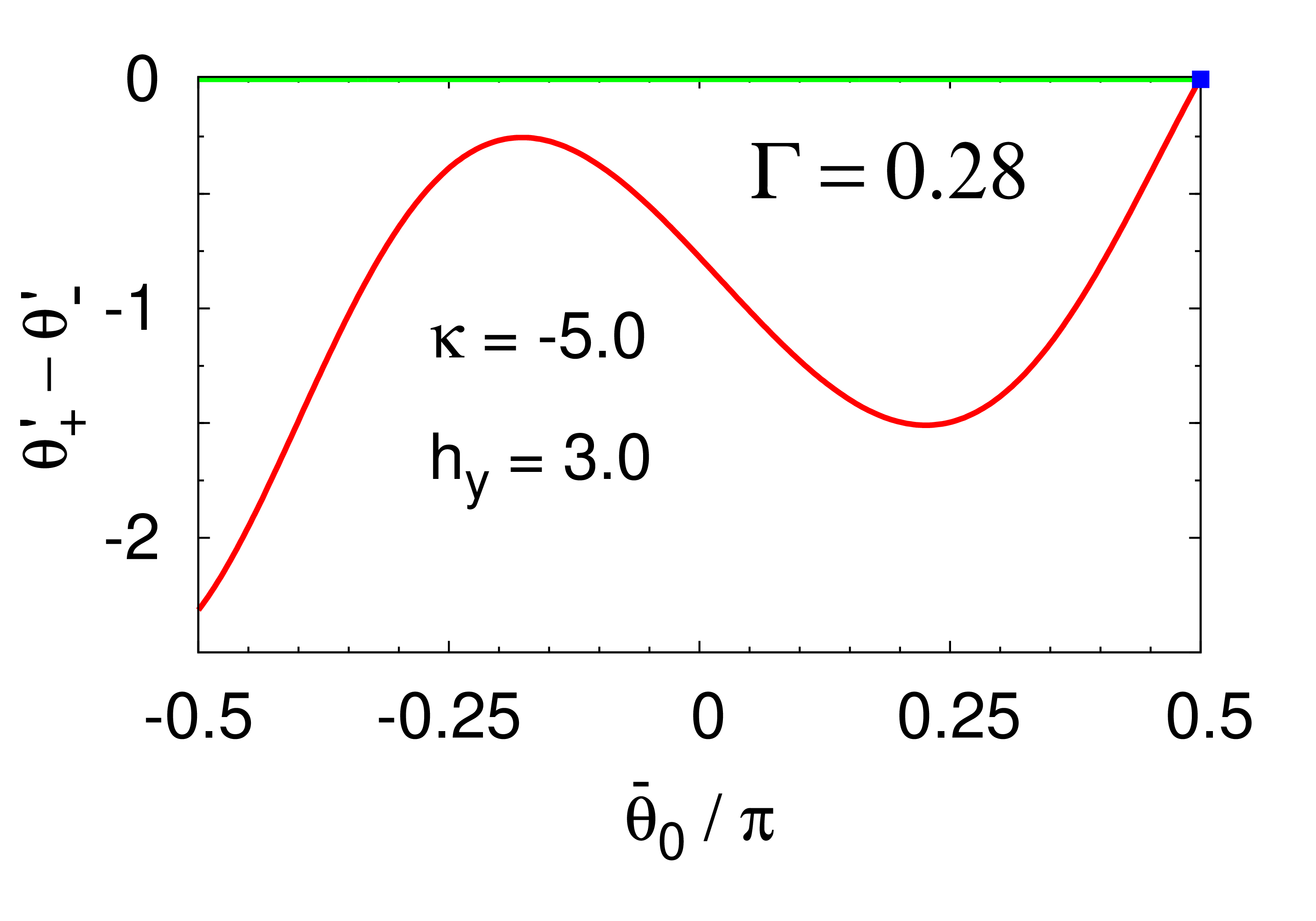}
\caption{
Matching at $w=0$ of the BVPs for $w\leq 0$ and $w\geq 0$, given by 
$\theta_+^\prime-\theta_-^\prime$, as a function of $\bar{\theta}_0$,
for $\Omega=0$, several values of $\Gamma$, $\kappa=-5.0$ and $h_y=1.0$
(top panels) and $h_y=3.0$ (bottom panels).
The blue filled circles and squares mark, respectively,
stable and unstable solutions of the BVP for $-\infty<w<\infty$.
The green lines show $\lim_{w\rightarrow 0^+}\varphi^\prime(w)-\lim_{w\rightarrow 0^-}\varphi^\prime(w)$,
which is zero to machine precision.
\label{fig:solBVP}}
\end{figure}

This BVP has no solution in general. 
To obtain a solution it is necessary to impose some relation between $\Omega$ and $\Gamma$.
To see why the BVP has no solution in general, let us analyze the general form of the
solution in the asymptotic region $w\rightarrow\pm\infty$.

In the $w\rightarrow -\infty$
region, $\varphi^\prime$ and $\bar{\theta}=\pi/2-\theta$ are exponentially small and
the linearized equations have the form
\begin{eqnarray}
  \bar{\theta}^{\prime\prime} &=& (h_y-\kappa)\bar{\theta} - \Omega\bar{\theta}^\prime
  - \Gamma\varphi^\prime,
\label{eq:theta_as_minus} \\
\varphi^{\prime\prime} &=& h_y\varphi + \Gamma\bar{\theta}^\prime - \Omega\varphi^\prime.
\label{eq:phi_as_minus}
\end{eqnarray}
The solution is 
$\bar{\theta}=\bar{\theta}_\nu\exp{\nu z}$ and
$\varphi=\varphi_\nu\exp{\nu z}$, with $\nu$ a solution of
\begin{equation}
(\nu^2+\Omega\nu-h_y+\kappa)(\nu^2+\Omega\nu-h_y)+\Gamma^2\nu^2 = 0.
\label{eq:sec}
\end{equation}
This equation has four solutions. For $\Omega=\Gamma=0$ they are
$\nu=\pm\sqrt{h_y}$ and $\nu=\pm\sqrt{h_y-\kappa}$. Hence, two values of $\nu$
are positive and two negatives. Therefore, at least for small $\Omega$ and $\Gamma$, two
values of $\nu$ will have positive real part and two negative real part.
Let us call $\nu_i$, with $i=1,\ldots,4$, the four solutions of~(\ref{eq:sec}),
with $i=1,2$ having positive real part and $i=3,4$ negative real part.
The general asymptotic solution as $w\rightarrow -\infty$ is
\begin{equation}
\begin{pmatrix}
\bar{\theta} \\ \varphi
\end{pmatrix}
=
\sum_{i=1}^4 a_i^{(-)}\mathrm{e}^{\nu_iw}
\begin{pmatrix}
u_i \\ v_i
\end{pmatrix}.
\label{eq:as_sol_minus}
\end{equation}
where the vector $(u_i,v_i)^T$ is a solution of
\begin{equation}
\begin{pmatrix}
\nu_i^2 + \Omega\nu_i -h_y+\kappa & -\Gamma\nu_i \\
\Gamma\nu_i & \nu_i^2 + \Omega\nu_i -h_y
\end{pmatrix}
\begin{pmatrix}
u_i \\ v_i
\end{pmatrix}
= 0.
\label{eq:algebraic_as}
\end{equation}
To satisfy the BCs for $z\rightarrow -\infty$ it is necessary (and sufficient) 
that $a_3^{(-)}=a_4^{(-)}=0$.

For $w\rightarrow +\infty$ the functions $\bar{\theta} = \pi/2-\theta$ and
$\bar{\varphi}=\chi (2\pi-\varphi)$ are exponentially small and the
linearized equations read
\begin{eqnarray}
\bar{\theta}^{\prime\prime} &=& (h_y-\kappa)\bar{\theta} 
- \Omega\bar{\theta}^\prime + \chi\Gamma\bar{\varphi}^\prime,
\label{eq:theta_as_plus} \\
\bar{\varphi}^{\prime\prime} &=& h_y\bar{\varphi} - \chi\Gamma\bar{\theta}^\prime
- \Omega\bar{\varphi}^\prime.
\label{eq:phi_as_plus}
\end{eqnarray}
Notice that these equations are obtained from Eqs. (\ref{eq:theta_as_minus}) and (\ref{eq:phi_as_minus}) just
replacing $\Gamma$ by $-\chi\Gamma$. Hence, the general solution is
\begin{equation}
\begin{pmatrix}
\bar{\theta} \\ \bar{\varphi}
\end{pmatrix}
=
\sum_{i=1}^4 a_i^{(+)} \mathrm{e}^{\nu_iw}
\begin{pmatrix}
u_i \\ v_i
\end{pmatrix},
\label{eq:as_sol_plus}
\end{equation}
where the $\nu_i$ are the solutions of~(\ref{eq:sec}) and the vectors
$(u_i,v_i)^T$ are the corresponding solution of~(\ref{eq:algebraic_as}) in which
$\Gamma$ has to be replaced by $-\chi\Gamma$.
The B.C. for $w\rightarrow +\infty$ require $a_1^{(+)}=a_2^{(+)}=0$.

In general, a BVP can have one, many, or no solution. An Initial Value Problem
(IVP), however, has one and only one solution. We may try to solve the BVP problem
as an IVP with initial conditions at $w=0$ given by
\begin{equation}
  \varphi(0)=\pi, \quad \varphi^\prime(0)=\varphi^\prime_{\mathrm{ini}}, \quad
  \theta(0) = \theta_{\mathrm{ini}},  \quad \theta^\prime(0) = \theta^\prime_{\mathrm{ini}}.
\end{equation}
The asymptotic solutions of the IVP as $w\rightarrow\pm\infty$ are given by Eqs.
(\ref{eq:as_sol_minus}) and (\ref{eq:as_sol_plus}), with coefficients $a_i^{(\pm)}$
that are functions of the initial conditions $\varphi^\prime_{\mathrm{ini}}$,
$\theta_{\mathrm{ini}}$, and
$\theta^\prime_{\mathrm{ini}}$, and of $\Omega$ and $\Gamma$. The solution of the IVP will be
a solution of the BVP if
\begin{equation}
  a_3^{(-)}=a_4^{(-)}=a_1^{(+)}=a_2^{(+)}=0.
\end{equation}
These four equations cannot be solved in general only with the three initial
conditions as variables. They may be satisfied, however, if we tune either $\Omega$, or
$\Gamma$, or both. The initial conditions, however, suffice to enforce
either $a_3^{(-)}=a_4^{(-)}=0$ or $a_1^{(+)}=a_2^{(+)}=0$. Therefore, the
BVPs restricted either to the positive $(-\infty<w\leq 0)$ or
negative $(0\leq w<+\infty)$ region, with the BCs
\begin{equation}
  \varphi(0)=\pi, \quad \theta(0) = \theta_{\mathrm{ini}},
  \quad \varphi(-\infty)=0, \quad \varphi(+\infty)=\chi 2\pi
  \quad \theta(\pm\infty) = \theta_{\mathrm{ini}}, \label{eq:BCs}
\end{equation}
have a solution, that can be obtained by solving the IVP and using $\varphi_{\mathrm{ini}}^\prime$
and $\theta^\prime_{\mathrm{ini}}$ to enforce the BCs on each side ($w<0$ or $w>0$). Of course,
the values of $\varphi_{\mathrm{ini}}^\prime$ and $\theta^\prime_{\mathrm{ini}}$ will be different
for each of the two BVPs.

We solved the BVP by splitting it into two pieces,
one for $w\leq 0$ and another one for $w\geq 0$, with the BCs of Eqs.~(\ref{eq:BCs}).
We solved them numerically, using a relaxation method. A solution of the complete BVP, for
$-\infty<w<\infty$, is obtained from the two restricted BVP if the derivatives
$\theta^\prime$ and $\varphi^\prime$ are continuous at $w=0$.
To satisfy these two conditions, we have $\theta_{\mathrm{ini}}$ at our disposal as a variable, and,
given that this is not enough, we need to tune also $\Omega$ and $\Gamma$.
It turns out that $\varphi^\prime$
is continuous at $w=0$ if and only if $\Omega=0$, whatever $\theta_{\mathrm{ini}}$ or $\Gamma$.
Therefore, we set $\Omega=0$ and use $\theta_{\mathrm{ini}}$ to enforce the continuity of
$\theta^\prime$ at $w=0$. The condition $\Omega=0$ determines the soliton velocity through
Eq.~(15) of the main text, which is
\begin{equation}
v = \frac{\beta}{\alpha} b_jj. \label{eq:v}
\end{equation}
Then, $\Gamma=(\beta/\alpha-1) b_jj/v_0$, which remains as a free parameter
controlled by the current density.

Let us write $\theta_{\mathrm{ini}} = \pi/2 - \bar{\theta}_0$. For $\Omega=0$ and given $\Gamma$,
we solved the two restricted BVP for a sufficiently dense mesh of $\bar{\theta}_0$
from $-\pi/2$ to $\pi/2$ (since the polar angle $\theta$ takes values between
$0$ and $\pi$). Defining $\theta_\pm^\prime=\lim_{w\rightarrow 0^\pm}\theta^\prime(w)$, we
obtain a solution of the complete BVP if $\theta_+^\prime-\theta_-^\prime=0$.
Fig.~\ref{fig:solBVP} shows $\theta_+^\prime-\theta_-^\prime$ as a function of
$\bar{\theta}_0$ for $\kappa=-5.0$ and the values of $h_y$ and $\Gamma$ displayed
in the legends. When several zeros of $\theta_+^\prime-\theta_-^\prime$ appear,
the steady solution corresponding to $\theta_0$ closest $0$ is stable and the other
unstable.

\begin{figure}[t!]
\centering
\includegraphics[width=0.32\textwidth]{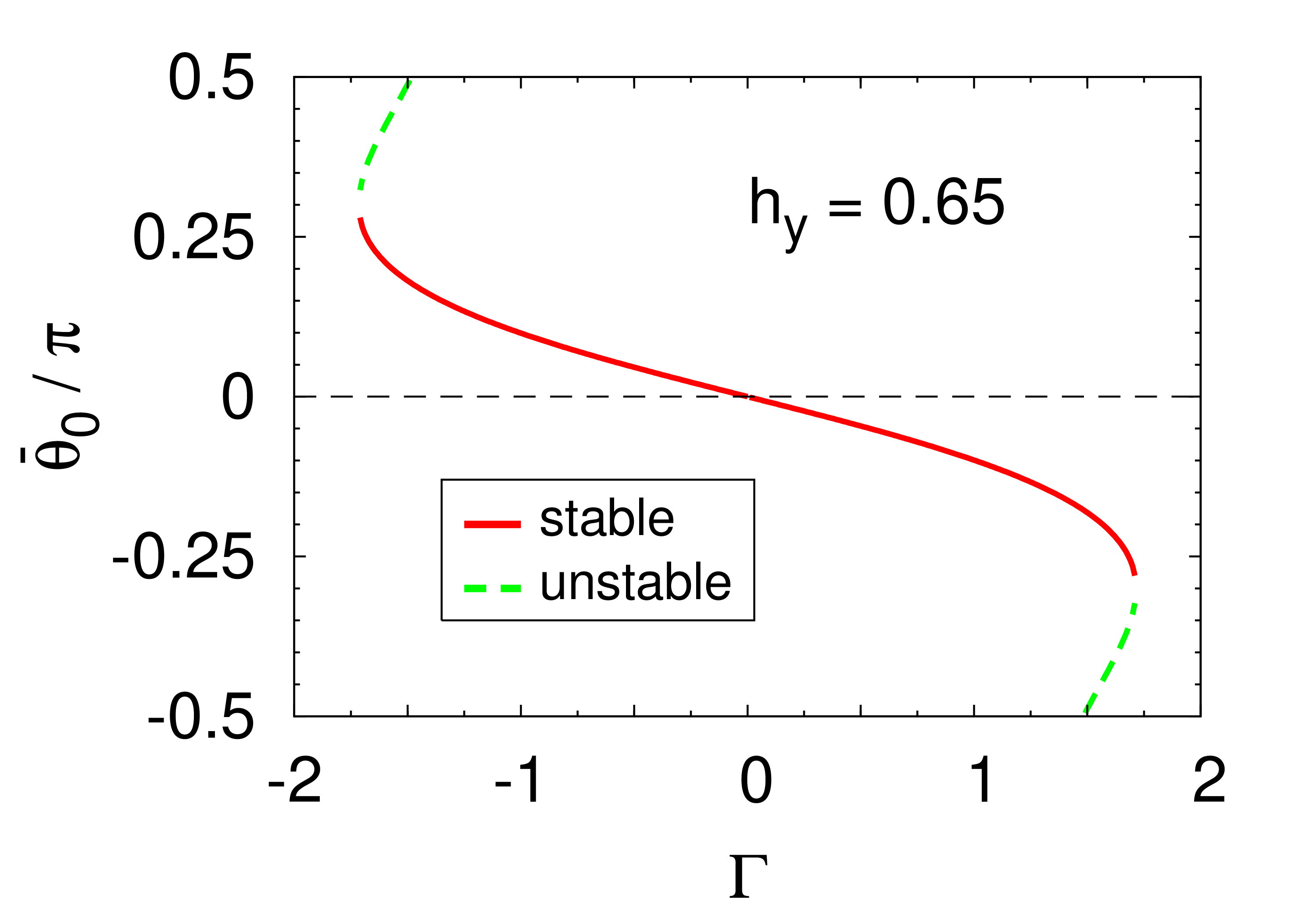}
\includegraphics[width=0.32\textwidth]{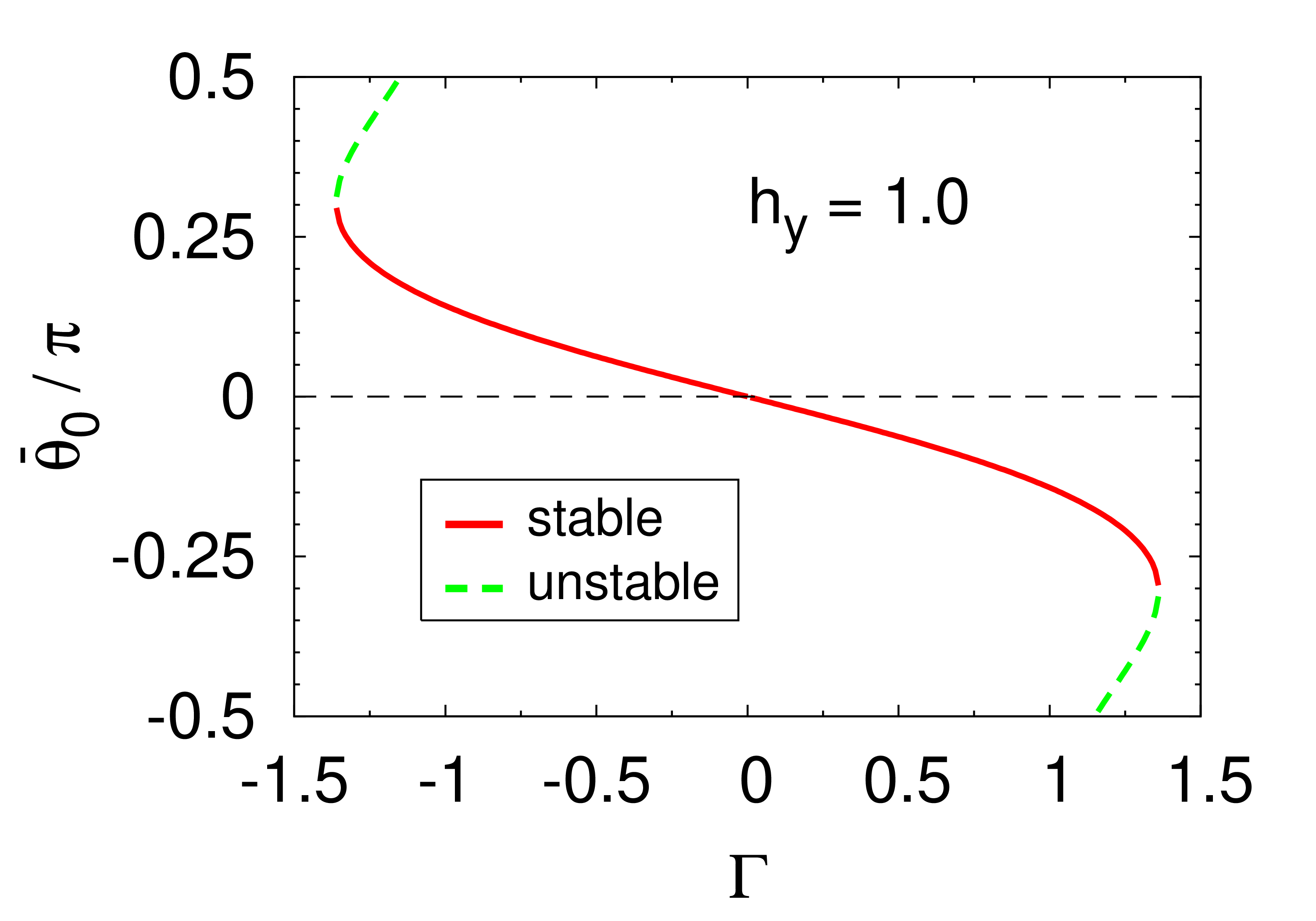}
\includegraphics[width=0.32\textwidth]{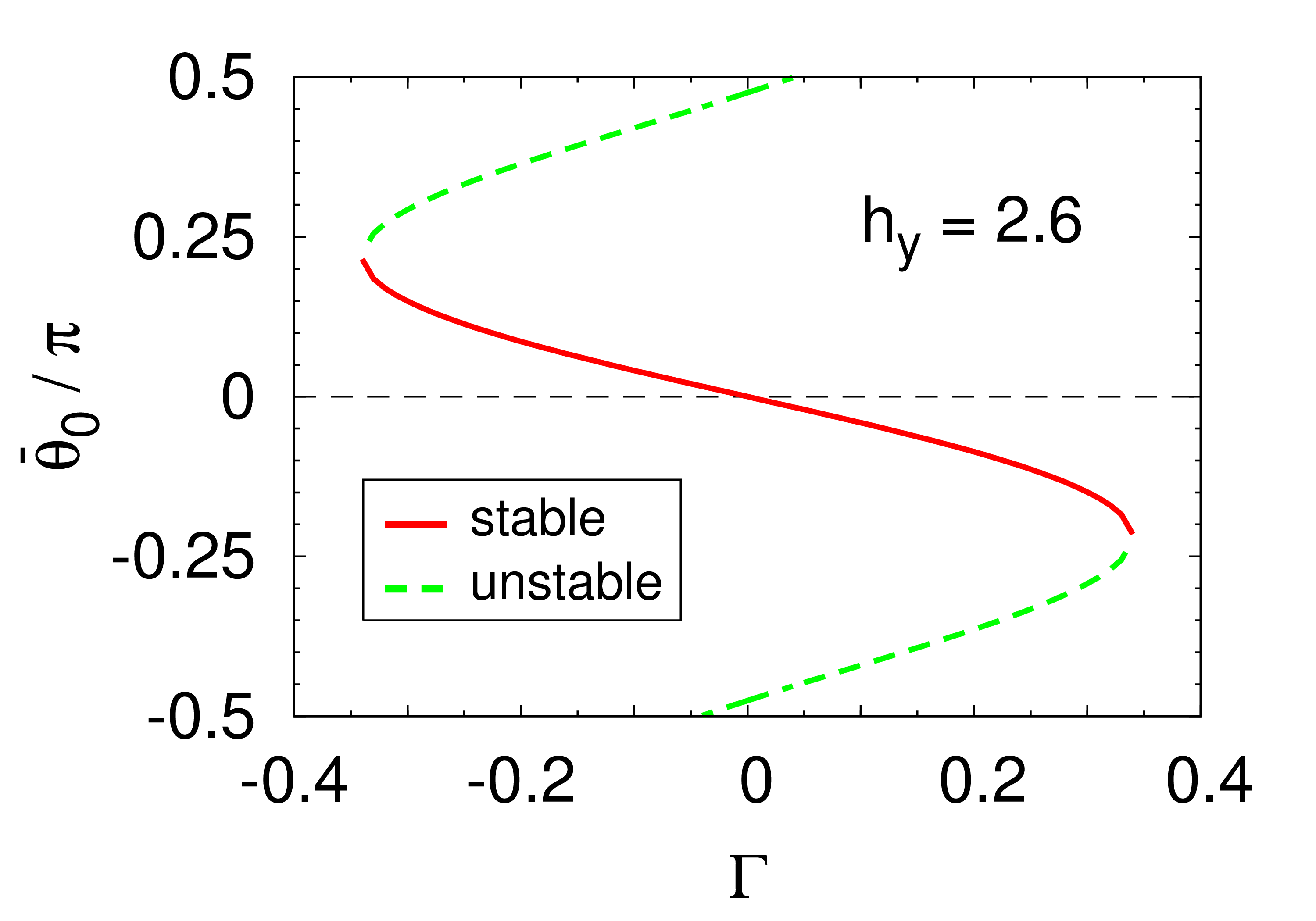} \\
\includegraphics[width=0.32\textwidth]{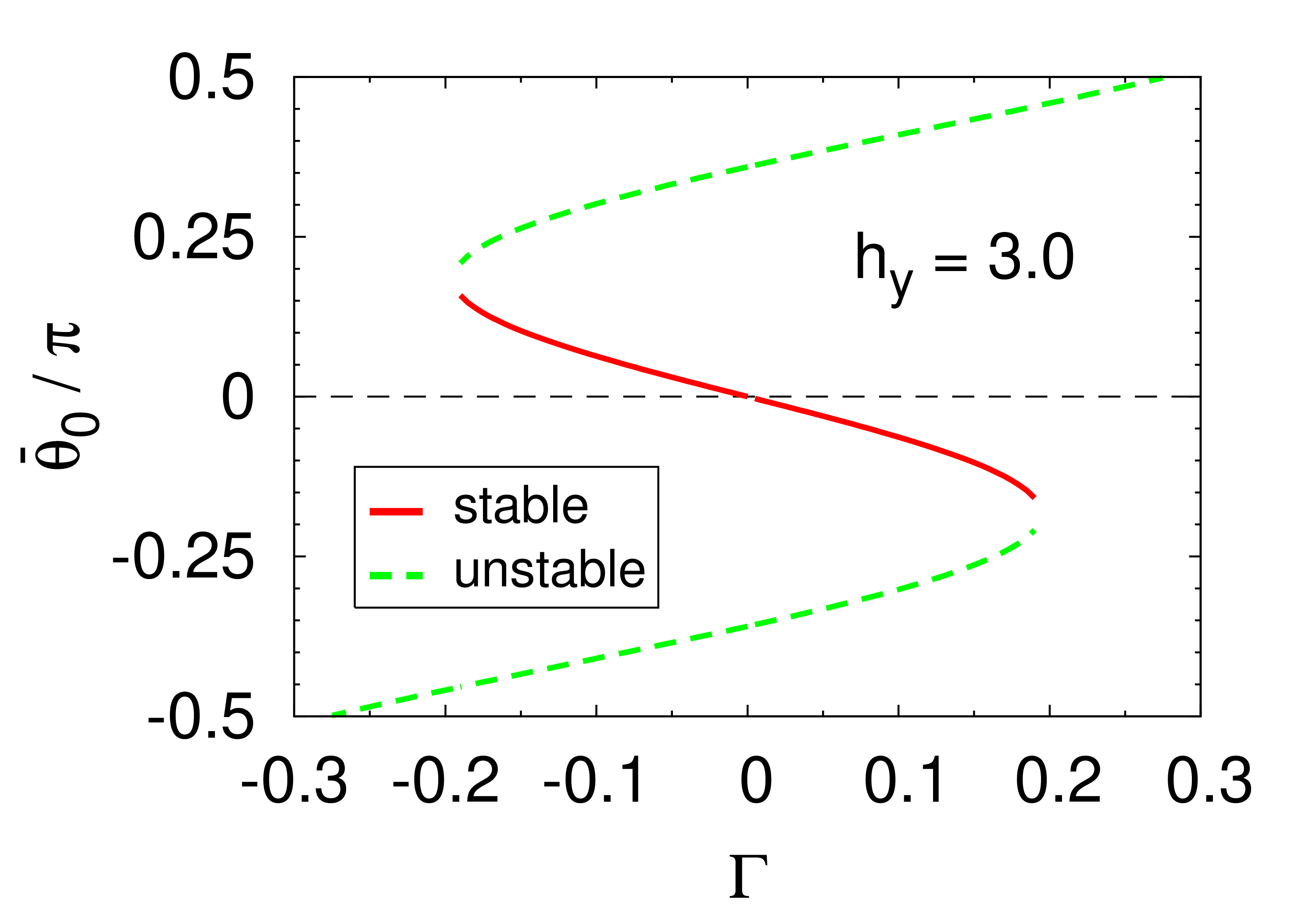}
\includegraphics[width=0.32\textwidth]{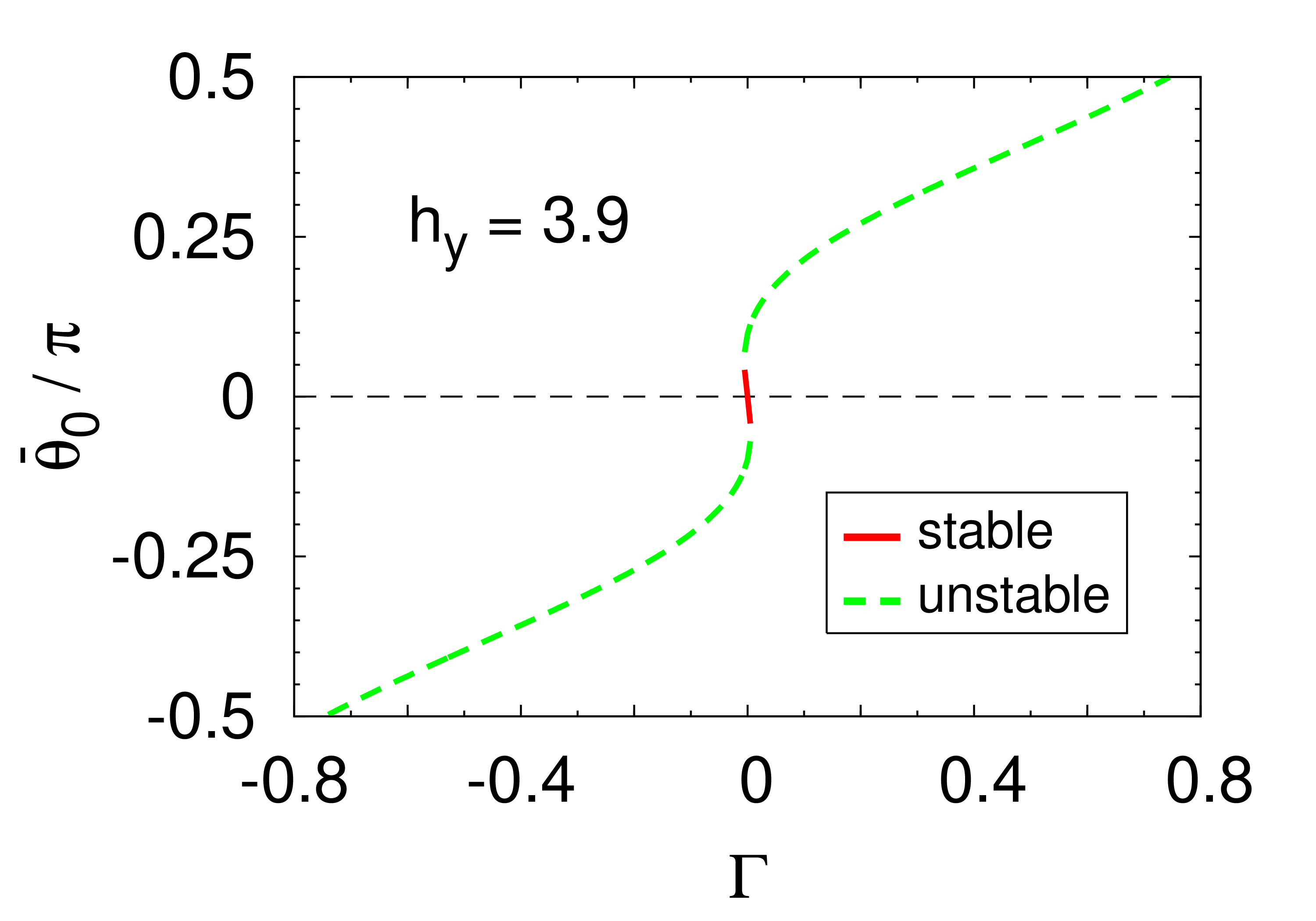}
\includegraphics[width=0.32\textwidth]{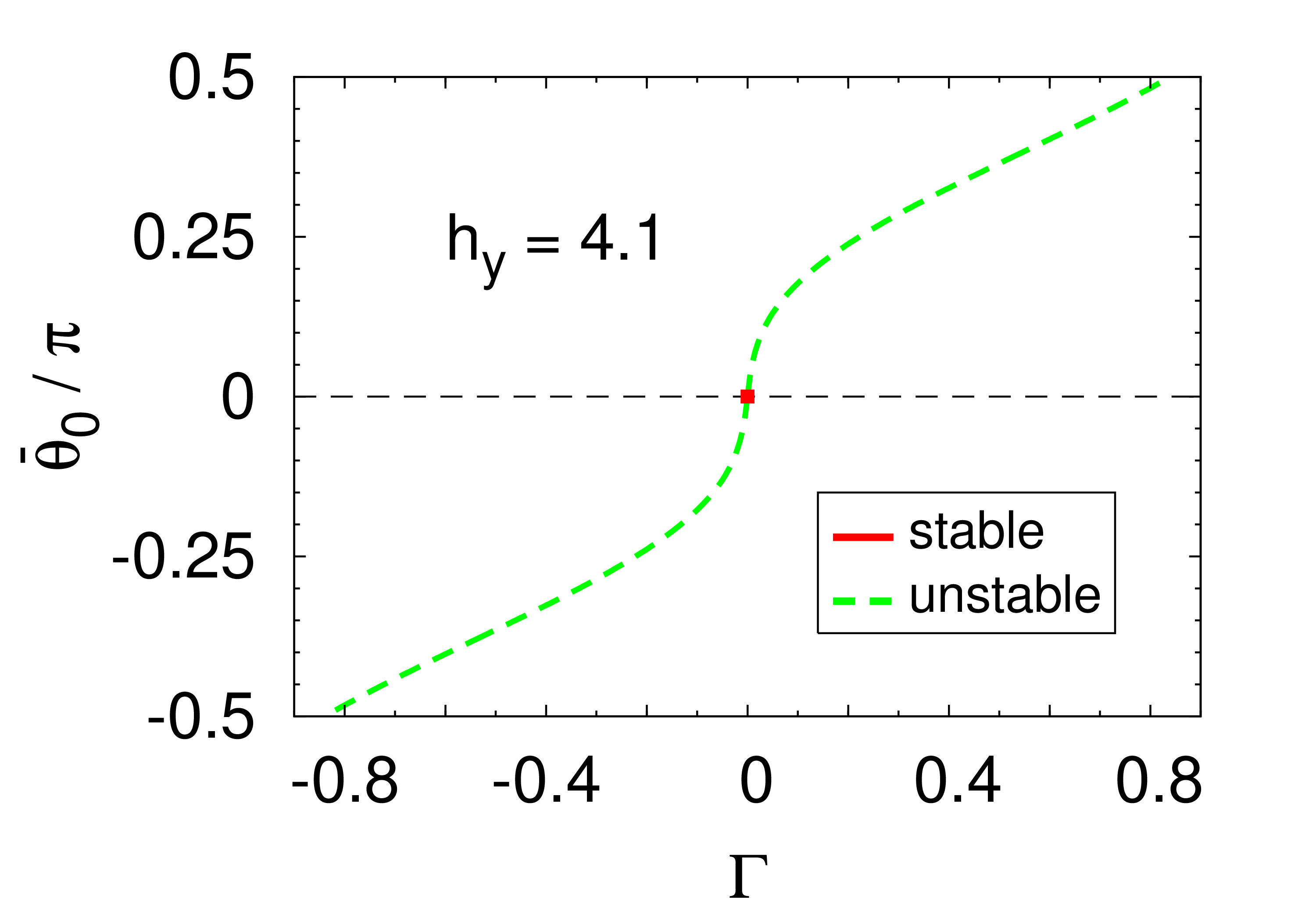}
\caption{Solutions of the BVP, parametrized by $\bar{\theta}_0$, as a function 
of $\Gamma$ for $\kappa=-5.0$ and the $h_y$ displayed in the legend.
The solid red lines are the stable solutions. The broken green lines are unstable
solutions. On the soliton stability boundary, $h_y\approx 4.1$, the stable solution is
reduced to the point $\Gamma=0$ and $\bar{\theta}_0=0$. 
\label{fig:sols}}
\end{figure}

Fig.~\ref{fig:sols} displays the solutions of the BVP, characterized by $\bar{\theta}_0$,
as a functionof $\Gamma$ in several cases. The solid red lines are the stable solutions
and the broken green lines the unstable solutions. The stable solution becomes unstable
at $\Gamma_c$, when it meets the unstable branch. 
For low $h_y$ (top panels) no steady solution exists for $|\Gamma|>|\Gamma_c|$.
For higher $h_y$ (bottom panels) unstable steady solutions exist for $|\Gamma|>|\Gamma_c|$.
As $h_y$ approaches from below the soliton stability boundary, which for $\kappa=-5.0$
corresponds to $h_y\approx 4.1$ [Fig.~2(a) of the main text], the stable solution branch
shrinks to zero, and, evidently, for higher values of $h_y$ even the static ($\Gamma=0$)
solution is unstable.

\section{\noindent S3. Numerical simulations details}

Numerical simulations of the magnetization dynamics were perfermed using the MuMax3 computational code~\cite{MuMax3, Exl15, Leliaert18}. The current release of the code includes implementations of the Dzyaloshinskii-Moriya interaction (DMI) in its bulk and interfacial form. Recently, so-called micromagnetic standard problems for ferromagnetic materials with DMI interactions has been proposed by Cort\'es-Ortu\~no \textit{et al} in Ref.~\cite{Cortes-Ortuno18}. Closely following the implementation of DMI interaction for crytals within the $D_{2d}$ symmetry class in~\cite{Cortes-Ortuno-github}, we have implemented the uniaxial DMI.
Notice that the coordinate system $(x,y,z)$ we are using, shown in Fig. 1 in the main text, is equivalent to $(y,z,x)$ in the MuMax3 implementation. In the present notation, the $D_{2d}$ DMI energy density term implemented in~\cite{Cortes-Ortuno18} has the form
\begin{equation}
 e_{\mathrm{DMI}} = D \hn \cdot \left( \partial_z \hn \times \hz + \partial_x \hn \times \hx \right) = D \left( n_x \partial_z n_y - n_y \partial_z n_x + n_y \partial_x n_z - n_z \partial_x n_y \right).
\end{equation}
For uniaxial helimagnets the DMI energy density can be expresses as
\begin{equation}
 e_{\mathrm{DMI}} = -D \hz \cdot \left( \hn \times \partial_z \hn \right) = -D \left( n_x \partial_z n_y - n_y \partial_z n_x \right).
\end{equation}
Here the DMI vector points in the $+z$ direction, favoring helical magnetization textures rotating counter-clockwise in the $x-y$ plane.
Accordingly, effective fields for the $D_{2d}$ DMI and uniaxial DMI in a two-dimensional system are
\begin{equation}
 \mu_0 H_{\mathrm{DMI}} = \frac{2D}{M_\mathrm{S}} \left[ \partial_z n_y \hx + \left( \partial_x n_z - \partial_z n_x \right) \hy - \partial_x n_y \hz \right],
\end{equation}
and
\begin{equation}
 \mu_0 H_{\mathrm{DMI}} = \frac{-2D}{M_\mathrm{S}} \left( \partial_z n_y \hx - \partial_z n_x \hy \right),
\end{equation}
respectively.
Therefore uniaxial DMI corresponds to the first term (in vectorial notation) of the $D_{2d}$ DMI with opposite sign. We have thus used the implementation of~\cite{Cortes-Ortuno-github}, which is based on the effective fields, without the terms associated to the second term of the energy density in vectorial notation.

\section{\noindent S4. Material parameters for \CrNbS}

We have chosen materials parameters for \CrNbS~based on the following criteria. In absence of a magnetic field the chiral soliton lattice (CSL) is the stable magnetization state. The period of the CSL is $L_0 = 48\,\mathrm{nm}$, corresponding to a wave number $q_0 = 2\pi/L_0 = 0.13\,\mathrm{nm^{-1}}$~\cite{Miyadai83, Togawa12}. In terms of micromagnetic parameters
\begin{equation}
 q_0 = \frac{D}{2A}.
\end{equation}
When an out-of-plane magnetic field $H_y$ is applied the chiral soliton lattice is destabilized at the perpendicular critical field $\mu_0 H_{yc} = 230\,\mathrm{mT}$ and the system goes to a uniform magnetization state in the out-of-plane direction~\cite{Togawa12}. Instead, if a magnetic field is applied in the $z$ direction, a conical state is formed below the parallel critial field $\mu_0 H_{zc} \approx 10 \mu_0 H_{yc} = 2.3\,\mathrm{T}$~\cite{Yonemura17}. Theoretically, when measuring magnetic fields in units of
\begin{equation}
 \mu_0 H_0 = \frac{D^2}{2 A M_{\mathrm{S}}},
\end{equation}
the perpendicular an parallel critical fields are $h_{yc} = \pi^2/16$ and $h_{zc} = 1-\kappa$, with
\begin{equation}
 \kappa = \frac{4 A K}{D^2}.
\end{equation}
Therefore, $\mu_0 H_0 = \mu_0 H_{yc}/h_{yc} = 370\,\mathrm{mT}$ and $h_{zc}/h_{yc} \approx 10$ for \CrNbS~implies
\begin{equation}
 \kappa \approx 1-\frac{10 \pi^2}{16} = -5.17.
\end{equation}
Finally, the saturation magnetization can be obtained as
\begin{equation}
 M_{\mathrm{S}} = \frac{g \mu_B S}{a^3} = 129\,\mathrm{kA/m},
\end{equation}
where $g=2$ is the Land\'{e} factor, $\mu_B$ the Bohr magneton, $S=3/2$ is the spin modulus and $a=0.6\,\mathrm{nm}$ is the lattice constant.

The micromagnetic parameters $A$, $K$ and $D$ can thus be obtained using $q_0 = 0.13\,\mathrm{nm^{-1}}$, $\kappa = -5.17$ and $\mu_0 H_0 = 307\,\mathrm{mT}$ through
\begin{eqnarray}
 A = \frac{\mu_0 H_0 M_{\mathrm{S}}}{2 q_0^2} = 1.42\,\mathrm{pJ/m}, \\
 K = \frac{\kappa \mu_0 H_0 M_{\mathrm{S}}}{2} = -124\,\mathrm{kJ/m^3}, \\
 D = \frac{\mu_0 H_0 M_{\mathrm{S}}}{q_0} = 369\,\mathrm{\mu J/m^2}.
\end{eqnarray}
In addition we assumed $\alpha = 0.01$ for the Gilbert damping parameter, $\beta = 0.02$ for the nonadiabaticity coefficient, and $P = 1$ for the polarization degree.

Finally, numerical simulations presented in this work have been performed without considering the demagnetizing field and with periodic boundary conditions in all directions.
A cell size of 1 nm $\times$ 1 nm $\times$ 1 nm were used in a system with dimensions $L_x = 2\,\mathrm{nm}$, $L_y= 1\,\mathrm{nm}$ and $L_z = 480\,\mathrm{nm}$.

\section{\noindent S5. Dynamics of the $\chi = -1$ chiral soliton}

When $D > 0$ the $\chi = +1$ soliton if favored against the one with the opposite chirality $\chi = -1$. Despite that, the $\chi = -1$ soliton can also be obtained as a metastable state of the system. Figure 2(a) of the main text shows that the region of metastability of the $\chi = -1$ soliton is reduced as compared with that of the $\chi = 1$ soliton. This means that a metastable $\chi = -1$ soliton can also be retained and forced under an applied current density. We have numerically tested this by driving the unfavored $\chi = -1$ soliton at $B_y = 0.2\,\mathrm{T}$.

\begin{figure}[t!]
\centering
\begin{subfigure}{0.46\textwidth}
\includegraphics[width=\textwidth]{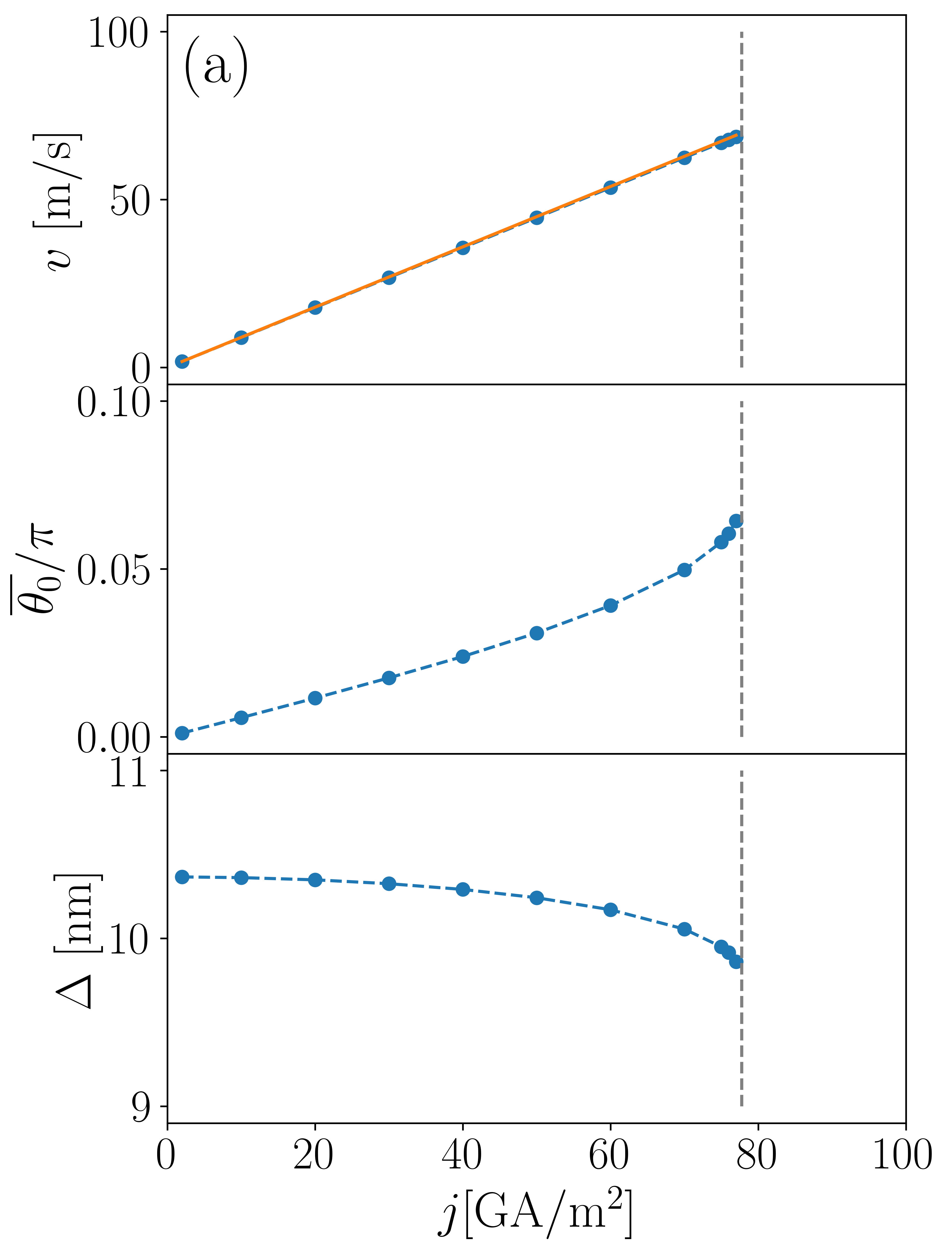}
\end{subfigure}
\begin{subfigure}{0.44\textwidth}
\includegraphics[width=\textwidth]{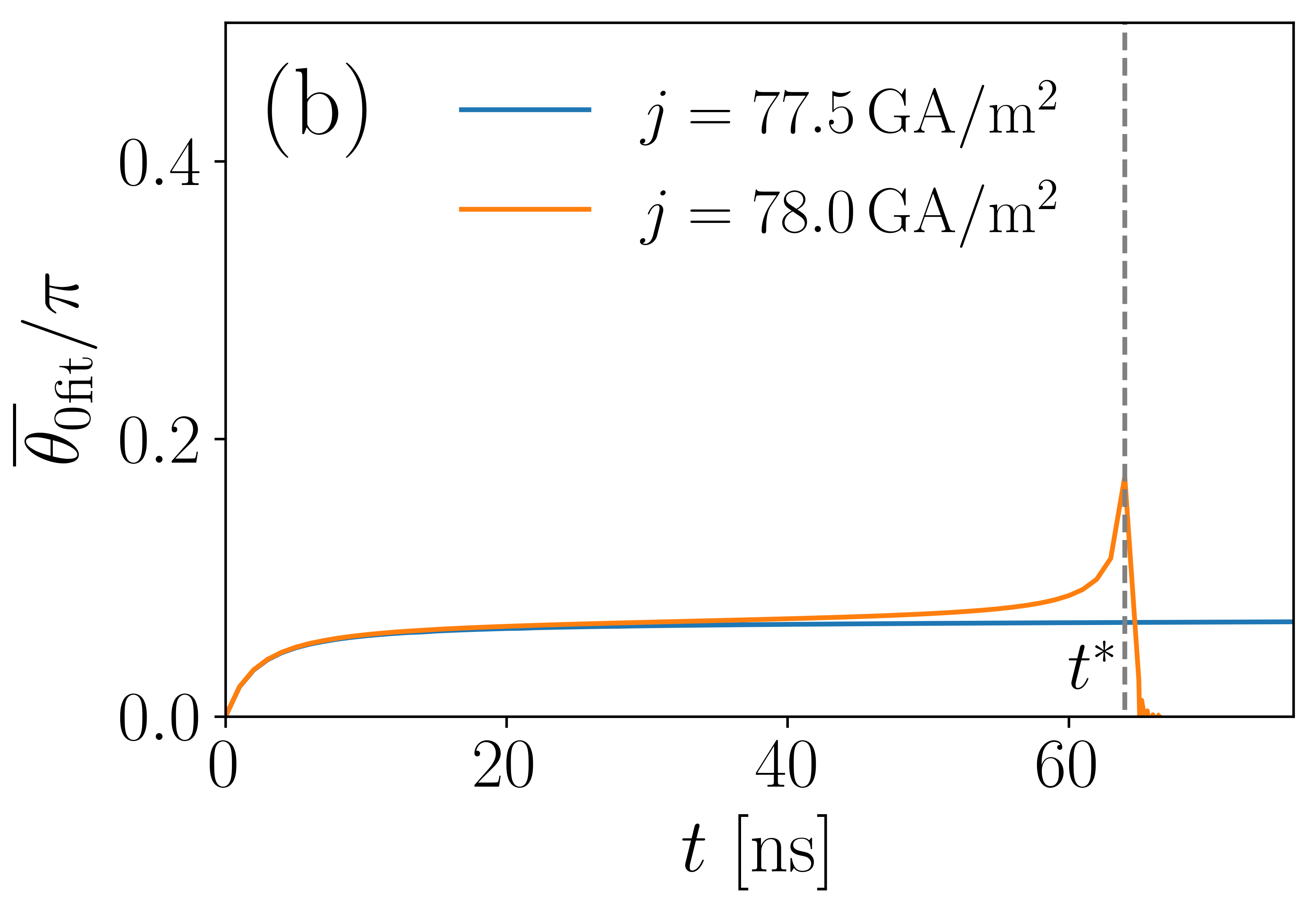}
\includegraphics[width=\textwidth]{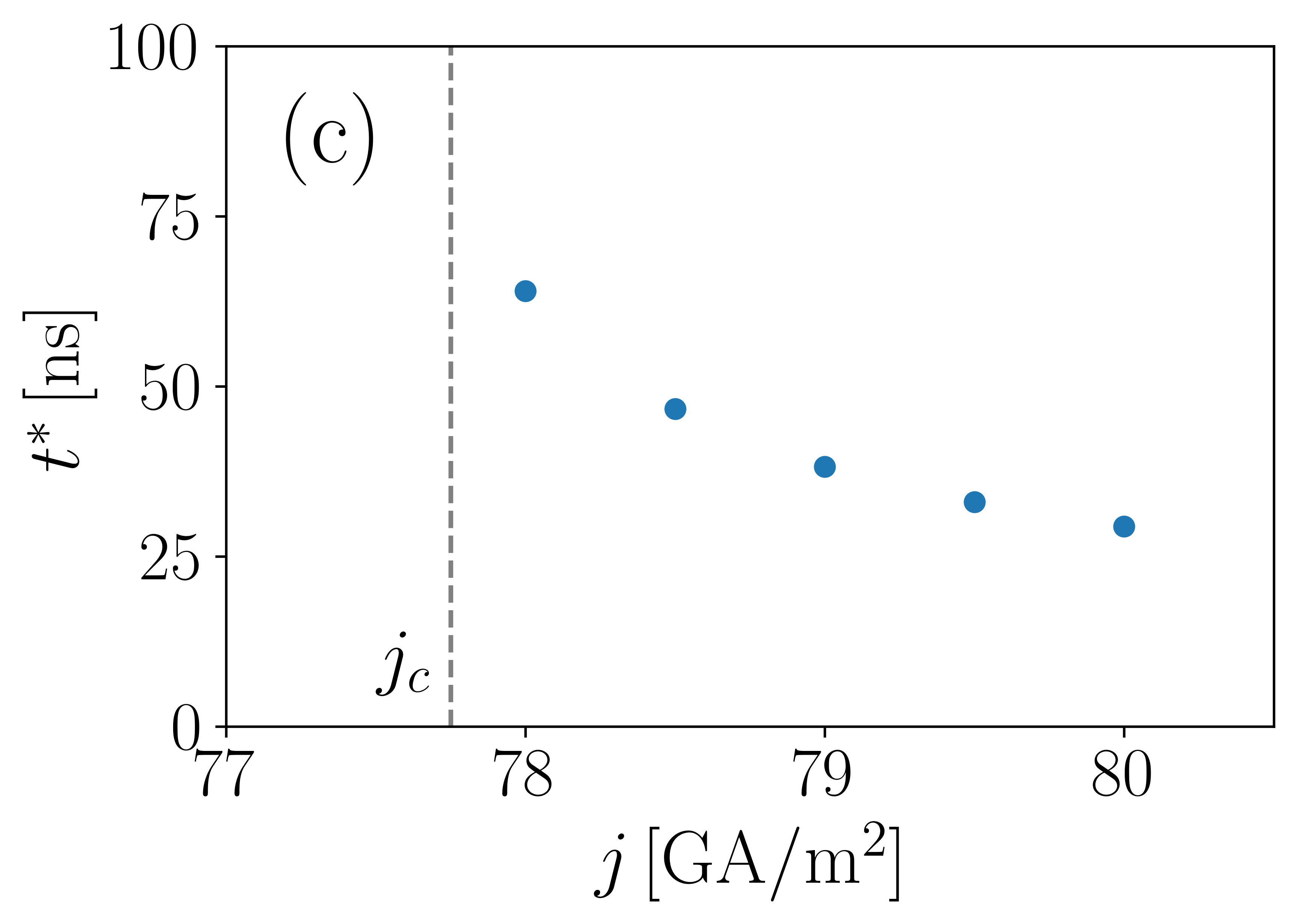}
\end{subfigure}
\caption{
(a) Velocity and soliton parameters as a function of the applied current density $j$ for the unfavored $\chi = -1$ soliton. The steady velocity increases linearly with the current, with the same mobility $m = (\beta/\alpha) b_j$ as for the $\chi = +1$ soliton, as indicated by the continuous line (top panel).
The tilt of the magnetization in the $z$ direction $\bar{\theta}_0$ is shown in the middle panel, presenting a noticeable increase when reaching $j_c = 77.75\,\mathrm{GA/m^2}$ (vertical lines).
Soliton width $\Delta$ decreases with the current density, as shown in the bottom panel.
(b) Evolution with time of the tilt angle $\bar{\theta}_0$ around the critical current $j_c$. The vertical line indicates the value of $t^*$ for $j= 78.0\,\mathrm{GA/m^2}$, beyond which the magnetization in the center of the soliton abruptly goes to the $y$ direction.
(c) Dependence on the current density of the instability time $t^*(j)$, showing how it seems to diverge when approaching $j_c = 77.75\,\mathrm{GA/m^2}$ from above.
\label{fig:chineg}}
\end{figure}

Figure~\ref{fig:chineg}(a) shows the evolution the velocity of the chiral soliton with the current density (upper paner). The mobility is $m = (\beta/\alpha) b_j$ and does not depend on the form or sign of the soliton. The tilt angle in the $z$ direction, $\bar{\theta}_0$, is shown in the middle panel of Fig.~\ref{fig:chineg}(a) while the soliton width $\Delta$ is presented in the bottom panel. A noticeable change is observed in both quantities when approaching the critical curren density $j_c = 77.75\,\mathrm{GA/m^2}$, indicated by vertical lines. The evolution of the tilt angle $\bar{\theta}_0$ around the critical current density is presented in Fig.~\ref{fig:chineg}(b), with $t^*$ indicating the time beyond which the single soliton is destroyed. Finally, Fig.~\ref{fig:chineg}(c) shows how $t^*$ increases when $j_c$ is approached from above.


\bibliography{references}